\documentclass[12pt, draftclsnofoot, onecolumn]{IEEEtran}
\usepackage{amsmath}
\usepackage{amssymb}
\usepackage{amsfonts}
\usepackage{algorithm}
\usepackage{algorithmic}
\usepackage{bbm}
\usepackage{booktabs}
\newcommand{\Rmnum}[1]{\expandafter\@slowromancap\romannumeral #1@}
\usepackage{graphicx}
\usepackage{epsfig}
\usepackage{subcaption}
\usepackage{citesort}
\usepackage{xcolor}
\usepackage{enumerate}
\usepackage{extarrows}
\makeatletter
\def\ScaleIfNeeded{%
\ifdim\Gin@nat@width>\linewidth \linewidth \else \Gin@nat@width
\fi } 
\renewcommand{\@IEEEsectpunct}{\ \, }% Modified from {:\ \,}
\makeatother

\begin{document}

\title{Joint Task Assignment and Resource Allocation for D2D-Enabled Mobile-Edge Computing}

\author{Hong Xing, Liang Liu, Jie Xu, and Arumugam Nallanathan
%\thanks{Manuscript received August 07, 2017; revised December 01, 2017; accepted January 20, 2018. This work was supported by the Engineering and Physical Sciences Research Council (EPSRC) of UK under Grant EP/N005651/2. Part of this paper was accepted by IEEE Wireless Communications and Networking Conference (WCNC), Barcelona, Spain, April, 2018 \cite{xing2017WCNC}. The associated editor coordinating the review of this paper and approving it for publication was Prof. Derrick Wing Kwan Ng.}
\thanks{Part of this paper has been presented at the IEEE International Conference on Communications (ICC), Kansas City, MO, USA, May, 2018 \cite{xing2018fixed}.}
%\thanks{H. Xing is with the College of Information Engineering, Shenzhen University, Shenzhen, 518060, China (e-mail: helen881005@gmail.com).}
\thanks{H. Xing and A. Nallanathan are with the School of Electronic Engineering and Computer Science, Queen Mary University of London, London, E1 4NS, U.K. (e-mails: h.xing@qmul.ac.uk, nallanathan@ieee.org).}
\thanks{L. Liu is with the Department of Electrical and Computer Engineering, National University of Singapore, Singapore (e-mail: eleliu@nus.edu.sg).}
\thanks{J. Xu is with the School of Information Engineering, Guangdong University of Technology, Guangzhou, 510006, China (e-mail: jiexu@gdut.edu.cn). J. Xu is also the corresponding author of this paper.}}

\maketitle
\vspace{-.2in}
\begin{abstract}
With the proliferation of computation-extensive and latency-critical applications in the 5G and beyond networks, mobile-edge computing (MEC) or fog computing, which provides cloud-like computation and/or storage capabilities at the network edge, is envisioned to reduce computation latency as well as to conserve energy for wireless devices (WDs). This paper studies a novel device-to-device (D2D)-enabled multi-helper MEC system, in which a local user solicits its nearby WDs serving as helpers for cooperative computation. We assume a time division multiple access (TDMA) transmission protocol, under which the local user offloads the tasks to multiple helpers and downloads the results from them over orthogonal pre-scheduled time slots. Under this setup, we minimize the computation latency by optimizing the local user's task assignment jointly with the time and rate for task offloading and results downloading, as well as the computation frequency for task execution, subject to individual energy and computation capacity constraints at the local user and the helpers. However, the formulated problem is a mixed-integer non-linear program (MINLP) that is difficult to solve. To tackle this challenge, we propose an efficient algorithm by first relaxing the original problem into a convex one, and then constructing a suboptimal task assignment solution based on the obtained optimal one. Next, we consider a benchmark scheme that endows the WDs with their maximum computation capacities. To further reduce the implementation complexity, we also develop a heuristic scheme based on the greedy task assignment. Finally, numerical results validate the effectiveness of our proposed algorithm, as compared against the heuristic scheme and other benchmark ones without either joint optimization of radio and computation resources or task assignment design. 

\end{abstract}

\IEEEpeerreviewmaketitle
\newtheorem{definition}{\underline{Definition}}[section]
\newtheorem{fact}{Fact}
\newtheorem{assumption}{Assumption}
\newtheorem{theorem}{\underline{Theorem}}[section]
\newtheorem{lemma}{\underline{Lemma}}[section]
\newtheorem{proposition}{\underline{Proposition}}[section]
\newtheorem{corollary}[proposition]{\underline{Corollary}}
\newtheorem{example}{\underline{Example}}[section]
\newtheorem{remark}{\underline{Remark}}[section]
\newcommand{\mv}[1]{\mbox{\boldmath{$ #1 $}}}
\newcommand{\mb}[1]{\mathbb{#1}}
\newcommand{\Myfrac}[2]{\ensuremath{#1\mathord{\left/\right.\kern-\nulldelimiterspace}#2}}

\section{Introduction}
It is envisioned that by the year of 2020, around $50$ billions of interconnected Internet of things (IoT) devices will surge in wireless networks, featuring new applications such as video stream analysis, augmented reality, and autonomous driving. The unprecedented growth of these latency-critical services requires extensive real-time computation, which, however, is hardly affordable by conventional mobile-cloud computing (MCC) systems that usually deploy cloud servers far away from end users \cite{Barbarossa2014survey}. Compared with MCC, mobile-edge computing (MEC) endows cloud-computing capabilities within the radio access network (RAN) such that the users can offload computation tasks to edge servers in their proximity for remote execution and then collect the results from them with enhanced energy efficiency and reduced latency (see \cite{mao2017survey} and the references therein). 
%Complying with the concept of pulling cloud-like computing and storage resources to the very edge of the network, {\em fog computing} has been recognized as a generalized form of MEC with the definition of edge devices extended from base stations (BSs) to a broad range of {\em cloudlets}. Moreover, MEC is also integrated as an essential functionality of the fog radio access network (FRAN) architecture that has the potent to provide ultra low-latency communications and computation for the 5G systems and beyond \cite{Simeone2016Mag}. 
Meanwhile, in industry, technical specifications and standard regulations are also being developed by, e.g., the European Telecommunications Standard Institute (ETSI) \cite{ETSI14}, Cisco \cite{cisco15}, and the 3rd Generation Partnership Project (3GPP) \cite{3GPP17}.

%Cisco described {\em fog computing}\footnote{Since MEC and fog computing \cite{Shih2017network} share lots of concepts in common, we use these two terms interchangeably throughout the paper.} as a new model for analysing and acting on IoT data in \cite{cisco15}. Very recently, functionality support for MEC has also been identified by the 3rd Generation Partnership Project (3GPP) \cite{3GPP17}. 
%Microsoft's recently developed algorithms tailored for resource-scarce devices \cite{Microsoft17}, also pursue to fulfil technical requirements for MEC.

%Joint computation and wireless resource allocation (single-server single-user multi-user single-server,partial offloading -> binary offloading)2
Various efforts have been dedicated to addressing technical challenges against different computation offloading models. In general, computation task models fall into two major categories, namely the partial offloading and the binary offloading. Tasks that cannot be partitioned for execution belong to the former category, while the latter category supports fine-grained task partitions. On the other hand, there are various types of MEC system architectures, such as single-user single-server \cite{Mao2017scheduling,Munoz2015EE,Cao18MEC}, multi-user single-server \cite{Sardellitti2015multicell,Chen2018MCC,Bi2018MEC,You2017offloading,Wang2018NOMA,Osvaldo2017AR,He2018MEC,Wang2018WP,Xu18PLS}, as well as single/multi-user multi-server  \cite{Shi2012Serendipity,Chen2017massiveD2D,Kao2017Hermes,Tran2017MEC}.  
To achieve satisfactory trade-offs between energy consumption and computing latency under different setups, it is critical to jointly optimize the radio and computation resources. Under a single-user single-server setup, \cite{Mao2017scheduling} jointly optimized the task offloading scheduling and the transmitting power allocation to minimize the weighted sum of the execution delay and the system energy consumption. The optimal communication strategy as well as computational load distribution was obtained in \cite{Munoz2015EE} in terms of the trade-offs of energy consumption versus latency for a multi-antenna single-user single-server system. In a single-user single-helper single-server system, \cite{Cao18MEC} proposed a novel user cooperation in both computation and communication to improve the energy efficiency for latency-constrained computation. In a multiple-input multiple-output (MIMO) multicell system, \cite{Sardellitti2015multicell} jointly optimized the precoding matrices of multiple wireless devices (WDs) and the CPU frequency assigned to each device with fixed binary offloading decisions, in order to minimize the overall users' energy consumption. A multi-user MCC system with a computing access point (CAP), which can serve as both a network gateway connecting to the cloud and an edge cloudlet, was studied in \cite{Chen2018MCC} to find the binary offloading decisions. Joint optimization of (partial) offloaded data length and offloading time/subcarrier allocation was studied in a multi-user single-server MEC system based on time-division multiple access (TDMA) and orthogonal frequency-division multiple access (OFDMA), respectively, in \cite{You2017offloading}. Furthermore, both binary and partial offloading were considered in a multi-user single-server MEC system exploiting multi-antenna non-orthogonal multiple access (NOMA) in \cite{Wang2018NOMA}. References \cite{Osvaldo2017AR} and \cite{He2018MEC} leveraged the inherent collaborative properties of augmented reality (AR) applications across multiple WDs to minimize the users' total energy expenditure.

%Task assignment (in multi-user multi-server scenario)
In the above works, the edge servers are mostly assumed to be one integrated server. However, considering multi-user multi-server systems where more than one CAPs are distributed over the network, it becomes non-trivial to answer the fundamental questions such as how to distribute the tasks among multiple servers, and how to schedule multiple tasks on one single server \cite{Shi2012Serendipity,Chen2017massiveD2D,Kao2017Hermes,Tran2017MEC,Alsalih2005energy-aware}. Computation resource sharing among WDs with intermittent connectivity was considered as early as in \cite{Shi2012Serendipity}, in which a greedy task dissemination algorithm was developed to minimize task completion time. A polynomial-time task assignment scheme for tasks with inter-dependency was developed in \cite{Kao2017Hermes} to achieve guaranteed latency-energy trade-offs. However, this line of works often assumed the communication conditions (e.g., transmission rate and multiple access schemes) and/or computation capacities (e.g., execution rate) to be fixed or estimable by some service profiling, but ignored potential performance improvement brought by dynamic management over such resources (e.g., transmitting power, bandwidth, and computation frequency).

In this paper, we study a device-to-device (D2D)-enabled multi-helper MEC system, in which a local user offloads a number of independent computation tasks to multiple nearby WDs serving as helpers, such as smart wearable devices, cellphones, tablets, and laptops, via direct D2D links.
{\color{black}The motivation for us to study efficient task assignment and wireless resource allocation algorithms to facilitate D2D-enabled MEC are two-fold. First, as WDs constantly improve their capabilities (e.g., battery capacity, processing speed, and spectral efficiency), heterogeneity of radio and computation resources among WDs can be exploited to support various demanding services while achieving mutual benefits \cite{Chen2017massiveD2D}. Second, with the proliferation of WDs, some of them may be prohibited from directly accessing to BSs. In these cases, they can entrust some virtual central controllers managed by network operators to collect their task features, and assist in D2D-enabled MEC by pooling and sharing the resources among each other.} 
Assuming that the tasks cannot be further partitioned, we consider a TDMA communication protocol, under which the local user offloads tasks to different helpers and downloads computation results from them over orthogonal pre-scheduled time slots. We aim for minimizing the overall latency subject to individual energy and computation capacity constraints at the local user and the helpers. 

%Related work and our novelty compared with them  
{\color{black}A task offloading framework, {\em D2D fogging}, was proposed in \cite{Chen16D2D}, where WDs could share the computation and communication resources among each other via the assistance of network operators, and dynamic task offloading decisions were made to minimize the time-averaged total energy consumption. From the perspective of system model, we employ different communication protocol from that in \cite{Chen16D2D}. We adopt a three-phase TDMA protocol, under which task assignment becomes a generally NP-hard problem because the tasks offloaded to different helpers (scheduled in different TDMA slots) are in couple with each other. By contrast, under OFDMA, all tasks can be executed independent of each other subject to a common deadline constraint as shown in \cite{Chen16D2D}. Furthermore, as in this paper, each WD is assumed to be assigned with more than one task, the efficient matching-based algorithm that forms the building block of the online task assignment scheme in \cite{Chen16D2D} cannot be applied any more.
It is also worth noting that the major difference between this paper and the earlier conference version \cite{xing2018fixed} is that instead of fixing the processing capacities, we consider controllable computation frequencies by exploiting dynamic voltage and frequency scaling (DVFS) \cite{Mach17survey} to achieve improved overall latency. 
To our best knowledge, this paper is among the earliest works investigating TDMA-based joint binary task offloading and wireless resource allocation for multiple tasks in a single-user multi-helper MEC system.
} 
%Furthermore, in terms of the multiple access scheme, instead of equally allocating the orthogonal resources to each task, we consider data transmission over variable-length TDMA slots. 
%Furthermore, the transmission delay incurred by results downloading are explicitly modelled and optimized as well in this work.

{\color{black}The contributions of our paper are summarized as follows. First, we transform the computation latency minimization problem with complex objective function into an equivalent one by investigating the optimal structure of the solution. Next, we jointly optimize the tasks assignment, the task offloading time/rate, the local and remote task execution time/computation frequency, and the results downloading time/rate, subject to individual energy and computation frequency constraints at the local user and the helpers. However, since the formulated problem is a mixed-integer non-linear program (MINLP) that is difficult to solve in general, we propose an efficient algorithm to obtain a high-quality sub-optimal solution by relaxing the binary task assignment variables into continuous ones, and then constructing a suboptimal task assignment solution based on the optimal one to the relaxed problem. Furthermore, to reduce the implementation complexity, we also provide fixed-frequency  task assignment and wireless resource allocation as a benchmark, and design a greedy task assignment based joint optimization algorithm. Finally, we evaluate the performance of the proposed convex-relaxation-based algorithm as compared against the heuristic one and other benchmark schemes without joint optimization of radio and computation resources or without task assignment design.} 

The remainder of this paper is organized as follows. The system model is presented in Section \ref{sec:System Model}. The joint task assignment and wireless resource allocation problem is formulated in Section \ref{sec:Problem Formulation}. The convex-relaxation-based joint task assignment and wireless resource allocation algorithm is proposed in Section \ref{sec:Joint Task Assignment and Wireless Resource Allocation}, while two low-complexity benchmark schemes are proposed in Section \ref{sec:Benchmark Schemes}. Numerical results are provided in Section \ref{sec:Numerical Results}, with concluding remarks drawn in Section \ref{sec:Conclusion}.

{\it Notation}---We use upper-case boldface letters for matrices and lower-case boldface ones for vectors. ``Independent and identically distributed'' is simplified as $i.i.d.$, and $\triangleq$ means ``denoted  by''. A circularly symmetric complex Gaussian (CSCG) distributed random variable (RV) $y$ with mean \(u\) and variance \(\sigma^2\) is denoted by \(y\sim\mathcal{CN}(u,\sigma^2)\). A continuous RV $z$ uniformly distributed  over \([a,b]\) is denoted by \(z\sim\mathcal{U}[a,b]\). \(\mb{R}^{M\times N}\) and \(\mb{R}^N\) stand for the sets of real matrices of dimension \(M\times N\) and real vectors of dimension $N$, respectively. The cardinality of a set is represented by \(\vert\cdot\vert\). In addition, \(\mathcal{P}(N)\) means an $N$-degree polynomial.

\section{System Model}\label{sec:System Model}
We consider a multi-user cooperative MEC system that consists of one local user, and $K$ nearby helpers denoted by the set \(\mathcal{K}=\{1,\ldots,K\}\), all equipped with single antenna. For convenience, we define the local user as the $(K+1)$-th WD. Suppose that the local user has $L$ independent tasks\footnote{\color{black}In this paper, we do not consider interdependency among tasks enabling data transmission from one helper to another as in \cite{Shi2012Serendipity,Kao2017Hermes}, since even under this simple task model, it becomes clear later that task assignment among multiple D2D helpers over pre-scheduled TDMA slots has already been very demanding to solve.} to be executed, denoted by the set \(\mathcal{L}=\{1,\ldots,L\}\), and the input/output data length of each task is denoted by \(T_l\)/\(R_l\) in bits. In the considered MEC system, each task can be either computed locally, or offloaded to one of the $K$ helpers for remote execution. Let \(\mv\Pi\in\mb{R}^{L\times (K+1)}\) denote the task assignment matrix, whose $(l,k)$-th entry, denoted by \(\pi(l,k)\in\{0,1\}\), \(l\in\mathcal{L}\), \(k\in\mathcal{K}\cup\{K+1\}\), is given by
\begin{align*}
	\pi(l,k)=\begin{cases}
		1, & \mbox{if the}\ l\mbox{th task is assigned to the}\  k\mbox{th WD,} \\
		0, & \mbox{otherwise.}
	\end{cases}
\end{align*}
Also, define $\mathcal{L}^{(k)}=\{l\in\mathcal{L}:\pi(l,k)=1\}$ as the set of tasks that are assigned to WD $k$, $k\in\mathcal{K}\cup\{K+1\}$. It is worthy of noting that we assume \(\vert \mathcal{L}^{(k)}\vert\ge 1\), \(k\in\mathcal{K}\cup\{K+1\}\). That is, each WD including the local user should be assigned with at least one task, i.e., \(L\ge K+1\)\footnote{\color{black}In practice, when \(L\le K\), a group of $K^\prime$ helpers are required to be selected {\em a priori} such that \(L\ge K^\prime+1\). However, detailed design regarding such selection mechanism is out of the discussion of this paper, and is left as our future work.}. Define by \(C_{l}\) in cycles the number of CPU cycles required for computing the \(l\)th task, $l\in\mathcal{L}$ \cite{Chen2017massiveD2D,Tran2017MEC}. Also, denote the CPU frequency in cycles per second (Hz) at the $k$th WD as $f_k$, \(k\in\mathcal{K}\cup\{K+1\}\).

\begin{figure}[htp]
	\centering
	\includegraphics[width=4.0in]{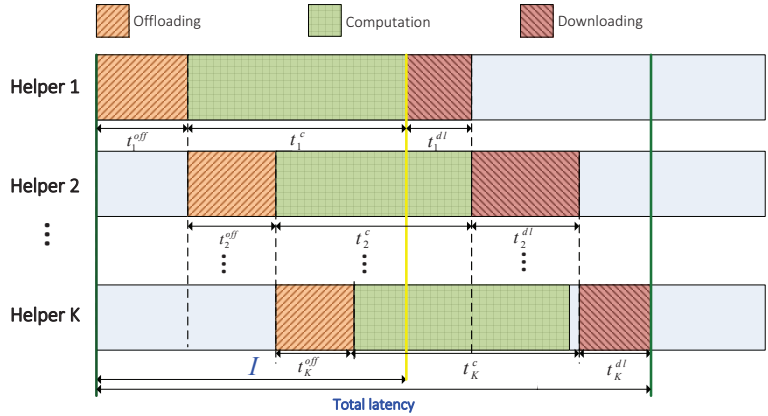}
	\caption{The TDMA-based frame structure for the proposed MEC protocol.}\label{fig:frame protocol}
	%\vspace{-0.2in}
\end{figure}

%\begin{figure}[htp]
%	\centering
%	\subfigure[A multi-user MEC system consisting of one local mobile device and several nearby wireless facilities equipped with MEC servers. \label{subfig:system illustration}]{\includegraphics[width=2.8in]{eps/system_model.eps}}
%	\subfigure[The TDMA-based frame protocol.\label{subfig:frame protocol}]{\includegraphics[width=3.2in]{eps/frame_protocol.eps}}
%	\caption{The system model.}\label{fig:system model}
%	%	\vspace{-0.2in}
%\end{figure}
\subsection{Local Computing}
The tasks in the set \(\mathcal{L}^{(K+1)}\) are executed locally with the local computation frequency in cycles per second given as \cite{Wang2018WP}
\begin{align}
	f_0=
	%\sum_{j=1}^{\sum_{l=1}^L\pi(l,K+1)C_lT_l}\Myfrac{1}{f_{0,j}}
	\frac{\sum_{l=1}^L\pi(l,K+1)C_{l}}{t_0^c}, \label{eq:local computation frequency}
\end{align} where \(t_0^c\) denotes the associated local computation time, and \(f_0\) is subject to the maximum frequency constraint, i.e., \(f_0\le f_0^{\max}\).
The corresponding computation energy consumed by the local user is given by \cite{Mach17survey}
\begin{align}
	E_0^c=
	%\sum_{j=1}^{\sum_{l=1}^L\pi(l,K+1)C_{l,0}T_l}\kappa_0f_{0,j}^2
	\kappa_0\sum_{l=1}^L\pi(l,K+1)C_lf_0^2, \label{eq:local computation energy}
\end{align}
where $\kappa_0$ is a constant denoting the effective capacitance coefficient that is decided by the chip architecture of the local user. Replacing \(f_0\) in \eqref{eq:local computation energy} with \eqref{eq:local computation frequency}, $E_0^c$ can thus be expressed in terms of $t_0^c$ as follows: 
\begin{align}
E_0^c=\kappa_0\frac{\left(\sum_{l=1}^L\pi(l,K+1)C_l\right)^3}{(t_0^c)^2}. \label{eq:local computation energy w.r.t time}  
\end{align}

\subsection{Remote Computing at Helpers}
The tasks assigned in \(\mathcal{L}^{(k)}\) is offloaded to the $k$th helper, \(k\in\mathcal{K}\), for remote execution. In this paper, we consider a three-phase TDMA communication protocol. As shown in Fig. \ref{fig:frame protocol}, the local user first offloads the tasks in the set $\mathcal{L}^{(k)}$ to the $k$th helper, \(k\in\mathcal{K}\), in a pre-scheduled order\footnote{\color{black}Since frequent change of TDMA scheduling policy incurs large amount of signalling overhead, we assume a fixed-order TDMA protocol in this paper, which is practically reasonable, and also commonly adopted in the literature, e.g., \cite{You2017offloading,Wang2018WP}.} via TDMA in the {\em task offloading} phase. Then the helpers execute their assigned computation tasks in the {\em task execution} phase. At last, in the {\em results downloading} phase, the helpers send the results back to the local user in the same order as in the task offloading phase via TDMA. Note that at each TDMA time slot during the task offloading phase, the local user only offloads tasks to one helper. Similarly, during the results downloading phase, only one helper can transmit over each time slot. In the following, we introduce the three-phase protocol in detail.
\subsubsection{Phase I: Task Offloading}
First, the tasks are offloaded to the helpers via TDMA. For simplicity, in this paper we assume that the local user offloads the tasks to the helpers with a fixed order of $1,2,\ldots,K$ as in Fig. \ref{fig:frame protocol}. In other words, the local user offloads tasks  $\mathcal{L}^{(1)}$ to the $1$st helper, then $\mathcal{L}^{(2)}$ to the $2$nd helper, until  $\mathcal{L}^{(K)}$ to the $K$th helper.

Let ${h}_k$ denote the channel power gain from the local user to the $k$th helper for offloading, \(k\in\mathcal{K}\). The achievable offloading rate (in bits per second) at the $k$th helper is given by
\begin{align}
	r_k^{off}=B\log_2\left(1+\frac{p_k^{off}{h}_k}{\sigma_k^2}\right),\label{eq:offloading rate}
\end{align}where \(B\) in Hz denotes the available transmission bandwidth, \(p_k^{off}\) is the transmitting power for offloading tasks to the $k$th helper, and $\sigma_k^2$ is the power of additive white Gaussian noise (AWGN) at the $k$th helper. Then, the time spent in offloading tasks to the $k$th helper is given by
%\footnote{If no task is offloaded to node $k$, i.e., $\pi(l,k)=0$, $\forall l$, then the offloading rate \(r_k^{off}=0\), and we define $t_k^{off}=0$ and $p_k^{off}=0$ in this case.}
\begin{align}
	t_k^{off}=\frac{\sum_{l=1}^L\pi(l,k)T_l}{r_k^{off}}. \label{eq:offloading time}
\end{align}
According to (\ref{eq:offloading rate}) and (\ref{eq:offloading time}), \(p_k^{off}\) is expressed in terms of \(t_k^{off}\) as
\begin{align}
	p_k^{off}=\frac{1}{\bar h_k}f\left(\frac{\sum_{l=1}^L\pi(l,k)T_l}{t_k^{off}}\right), \label{eq:offloading power}
\end{align}where $\bar h_k={h}_k/\sigma_k^2$ is the normalized channel power gain, and \(f(x)\) is a function defined as \(f(x)\triangleq2^{\frac{x}{B}}-1\). The total energy consumed by the local user for offloading all the tasks in \(\mathcal{L}\) is thus expressed as
\begin{align}
	E_0^{off}
	=\sum_{k=1}^K\frac{1}{\bar h_k}f\left(\frac{\sum_{l=1}^L\pi(l,k)T_l}{t_k^{off}}\right)t_k^{off}.\label{eq:offloading commun. energy}
\end{align}

\subsubsection{Phase II: Task Execution}
After receiving the assigned tasks $\mathcal{L}^{(k)}$, \(k\in\mathcal{K}\), the $k$th helper proceeds with the computation frequency given by
\begin{align}
	f_k=\frac{\sum_{l=1}^L\pi(l,k)C_{l}}{t_k^c}, \label{eq:remote computation frequency}
\end{align} where \(t_k^c\)'s is the remote computation time spent by the $k$th helper. Similarly, helper $k$'s remote computing frequency given by \eqref{eq:remote computation frequency} is also constrained by its maximum frequency, i.e.,\(f_k\le f_k^{\max}\). In addition, its computation energy is expressed as 
\begin{align}
	E_k^c=\kappa_k\frac{\left(\sum_{l=1}^L\pi(l,k)C_l\right)^3}{(t_k^c)^2},\label{eq:remote computation energy}
\end{align}where \(\kappa_k\) is the corresponding capacitance constant of the $k$th helper.

\subsubsection{Phase III: Results Downloading}
After computing all the assigned tasks, the helpers begin transmitting the computation results back to the local user via TDMA. Similar to the task offloading phase, we assume that the helpers transmit their respective results in the fixed order of $1,\ldots,K$.
Let ${g}_k$ denote the channel power gain from helper $k$ to the local user for downloading. The achievable downloading rate from the $k$th helper is then given by
\begin{align}
	r_k^{dl}=B\log_2\left(1+\frac{p_k^{dl}{g}_k}{\sigma_0^2}\right),\label{eq:feedback rate}
\end{align}where $p_k^{dl}$ denotes the transmitting power of the $k$th helper, and $\sigma_0^2$ denotes the power of AWGN at the local user. The corresponding downloading time is thus given by
\begin{align}
	t_k^{dl}=\frac{\sum_{l=1}^L\pi(l,k)R_l}{r_k^{dl}}. \label{eq:downloading time}
\end{align}
Combining \eqref{eq:feedback rate} and \eqref{eq:downloading time}, the transmitting power of the $k$th helper is expressed as
\begin{align}
	p_k^{dl}=\frac{1}{\bar g_k}f\left(\frac{\sum_{l=1}^L\pi(l,k)R_l}{t_k^{dl}}\right),\label{eq:downloading power}
\end{align}where $\bar g_k={g}_k/\sigma_0^2$ denotes the normalized channel power gain from the $k$th helper to the local user. The communication energy consumed by the $k$th helper is thus given by
\begin{align}
	E_k^{dl}=\frac{1}{\bar g_k}f\left(\frac{\sum_{l=1}^L\pi(l,k)R_l}{t_k^{dl}}\right)t_k^{dl}.\label{eq:downloading commun. energy}
\end{align}

\subsection{Total Latency}
Since TDMA is used in both Phase I and Phase III, each helper has to wait until it is scheduled. Specifically, the first scheduled helper, i.e., helper $1$, can transmit its task results to the local user only when the following two conditions are satisfied: first, its computation has been completed; and second, task offloading from the local user to all of the $K$ helpers are completed such that the wireless channels begin available for data downloading. As a result, helper $1$ starts transmitting its results after a period of waiting time given by
\begin{align}
	I_1=\max\{ t_1^{off}+t_1^c,\sum_{k=1}^Kt_k^{off}\}, \label{eq:waiting time of helper 1}
\end{align} where \(t_1^c\) is the task execution time at helper $1$.

Moreover, for each of the other $K-1$ helpers, it can transmit the results to the local user only when: first, its computation has been completed; second, the $(k-1)$th helper scheduled preceding to it has finished transmitting. Consequently, denoting the waiting time for helper $k$ (\(k\ge 2\)) to transmit the results as $I_k$, $I_k$ is expressed as
\begin{align}
	I_k=\max\{\sum_{j=1}^kt_j^{off}+t_k^c,I_{k-1}+t_{k-1}^{dl}\}. \label{eq:waiting time of helper k>=2}
\end{align} Accordingly, the completion time for all the results to finish downloading is expressed as
\begin{align}
	T=I_K+t_K^{dl}. \label{eq:completion time}
\end{align}

To sum up, taking local computing into account as well, the total latency for executing all of the $L$ tasks is given by
\begin{align}
	T^{\rm total}=\max\{t_0^c,T\}. \label{eq:total latency}
\end{align}

%It is easily verified that when the optimal $T^{\rm total}$ yields \(t_0^{c}>I_1+\sum_{k=1}^Kt_k^{dl}\), it is always possible, without loss of optimality, for one helper to slow down its transmission such that \(I_1+\sum_{k=1}^Kt_k^{dl}=t_0^{c}\) with its communication energy saved (c.f.~\eqref{eq:downloading commun. energy}). Therefore, $T^{\rm total}$ is simplified as \(T^{\rm total}=I_1+\sum_{k=1}^Kt_k^{dl}\) subject to \(t_0^{c}\le I_1+\sum_{k=1}^Kt_k^{dl}\).

\section{Problem Formulation}\label{sec:Problem Formulation}
In this paper, we aim at minimizing the total latency for local/remote computing of all the tasks by optimizing the task assignment strategy ($\pi(l,k)$'s), the task offloading time ($t_k^{off}$'s), the task execution time ($t_k^c$'s), and the results downloading time ($t_k^{dl}$'s), subject to the individual energy and computation frequency constraints at the local user as well as the $K$ helpers. Specifically, we are interested in the following problem:
\begin{subequations}
\begin{align}
\mathrm{(P0)}:&~\mathop{\mathtt{Minimize}}_{\mv\Pi,\{t_k^{off},t_k^{dl},t_k^c\},t_0^c}~~~T^{\rm total}\notag\\
&~~~\mathtt {Subject \ to}\notag\\
&~~~\kappa_0\frac{\left(\sum_{l=1}^L\pi(l,K+1)C_l\right)^3}{(t_0^c)^2}+\sum_{k=1}^K\frac{1}{\bar h_k}f\left(\frac{\sum_{l=1}^L\pi(l,k)T_l}{t_k^{off}}\right)t_k^{off}\le E_0,\label{C:energy constraint at the source}\\
&~~~\kappa_k\frac{\left(\sum_{l=1}^L\pi(l,k)C_l\right)^3}{(t_k^c)^2}+\frac{1}{\bar g_k}f\left(\frac{\sum_{l=1}^L\pi(l,k)R_l}{t_k^{dl}}\right)t_k^{dl}\le E_k,\; \forall k\in\mathcal{K},\label{C:energy constraint at the kth helper}\\
&~~~\frac{\sum_{l=1}^L\pi(l,K+1)C_{l}}{f_0^{\max}}\le t_0^c,\label{C:max frequency at the source}\\ 
&~~~\frac{\sum_{l=1}^L\pi(l,k)C_{l}}{f_k^{\max}}\le t_k^c,\; \forall k\in\mathcal{K},\label{C:max frequency at the kth helper}\\
&~~~\sum_{k=1}^{K+1}\pi(l,k)= 1, \; \forall l\in\mathcal{L},\label{C:all tasks assigned constraint}\\
&~~~\sum_{l=1}^L\pi(l,k)\ge 1, \; \forall k\in\mathcal{K}\cup\{K+1\}, \label{C:all users assigned constraint}\\
&~~~\pi(l,k)\in\{0,1\},\; \forall l\in\mathcal{L},\, k\in\mathcal{K}\cup\{K+1\},\label{C:binary constraint}\\
&~~~t_k^{off}\ge 0,\, t_k^{dl}\ge 0, \; \forall k\in\mathcal{K}.\label{C:nonnegative time}
\end{align}
\end{subequations}
In the above problem, the objective function \(T^{\rm total}\) is given by \eqref{eq:total latency}. The constraints given by \eqref{C:energy constraint at the source} and \eqref{C:energy constraint at the kth helper} state that the total energy consumption of computation and transmission for the local user and the $k$th helper cannot exceed \(E_0\) and \(E_k\)'s, respectively. In \eqref{C:energy constraint at the source}, $E_0^c$ and $E_0^{off}$ are replaced with \eqref{eq:local computation energy} and \eqref{eq:offloading commun. energy}, respectively, while \eqref{C:energy constraint at the kth helper} is obtained by substituting \eqref{eq:remote computation energy} and \eqref{eq:downloading commun. energy} for \(E_k^c\)'s and \(E_k^{dl}\)'s, respectively. 
\eqref{C:max frequency at the source} and \eqref{C:max frequency at the kth helper} guarantee that the computation frequencies of the local users (c.f.~\eqref{eq:local computation frequency}) and the helpers (c.f.~\eqref{eq:remote computation frequency}) stay below their respective limits. \eqref{C:all tasks assigned constraint} guarantees that each task must be and only assigned to one WD; and \eqref{C:all users assigned constraint} ensures that each of the local user and the helpers is assigned with at least one task. Finally, \eqref{C:binary constraint} imposes the binary offloading constraints.

\subsection{Problem Reformulation}
Note that $T_{\rm total}$ (c.f.~\eqref{eq:total latency}) is a complicated function involving accumulative \(\max(\cdot)\) mainly due to the recursive expression of $I_k$ (c.f.~\eqref{eq:waiting time of helper k>=2}). Hence, to obtain an explicit objective function in terms of the optimization variables, we need to simplify $T_{\rm total}$ exploiting the following proposition.
\begin{proposition}
Problem (P0) can be recast into an equivalent problem as follows.
\begin{subequations}
\begin{align}
\mathrm{(P0\text{-}Eqv)}:&~\mathop{\mathtt{Minimize}}_{\mv\Pi,\{t_k^{off},t_k^{dl},t_k^c\},t_0^c}~~~t_1^{off}+t_1^c+\sum_{k=1}^Kt_k^{dl}\notag\\
&~~~\mathtt {Subject \ to}\ \ \eqref{C:energy constraint at the source}-\eqref{C:nonnegative time},\notag\\
&~~~\sum_{k=1}^Kt_k^{off}\le t_1^{off}+t_1^c, \label{C:min computing time constraint for helper 1}\\
&~~~t_k^c\le t_1^{off}+t_1^c+\sum_{j=1}^{k-1}t_j^{dl}-\sum_{j=1}^kt_j^{off}, \; \forall k\in\mathcal{K}\setminus \{1\}, \label{C:computation deadline constraint for helper k>=2}\\
&~~~t_0^{c}\le t_1^{off}+t_1^c+\sum_{k=1}^Kt_k^{dl}. \label{C:computation deadline constraint for the local user}
\end{align} 
\end{subequations}\label{prop:simplification of T^total}
\end{proposition}
\begin{IEEEproof}
{\color{black}A brief idea of the proof is given as follows. To remove \(\max(\cdot)\) in \(I_k\)'s, we first need to narrow down from different cases leveraging the property of the optimal solution. Then, based on the simplified case, we recursively derive \(I_k\)'s for \(k\ge 2\). Finally, we arrive at a clear objective function of (P1-Eqv) subject to all the optimality conditions given by \eqref{C:min computing time constraint for helper 1}-\eqref{C:computation deadline constraint for the local user}.} Please refer to Appendix~\ref{appendix:proof of simplification of T^total} for the proof in detail.
\end{IEEEproof}	
%where \eqref{C:expanded energy constraint at the source} (\eqref{C:expanded energy constraint at the kth helper}) is obtained by replacing \(E_0^c\) (\(E_k^c\)'s) and \(E_0^{off}\) (\(E_k^{dl}\)'s) in \eqref{C:energy constraint at the source} (\eqref{C:energy constraint at the kth helper}) with \eqref{eq:local computation energy} (\eqref{eq:remote computation energy}) and \eqref{eq:offloading commun. energy} (\eqref{eq:downloading commun. energy}), respectively, and \eqref{C:min computing time at the source} (\eqref{C:min computing time at the kth helper}) is as a result of substituting  \eqref{eq:local computation frequency} (\eqref{eq:remote computation frequency}) for \(f_0\) (\(f_k\)'s). 
\subsection{Suboptimal Design}
The transformed problem (P0-Eqv) is seen as an MINLP due to the integer constraints given by \eqref{C:binary constraint}, and is thus in general NP-hard. Although the optimal solution to (P0-Eqv) can be obtained by exhaustive search, it is computationally too expensive (approx. \(O((K+1)^L)\) times of search) to implement in practice. Therefore, we solicit two approaches for suboptimal solution to (P0-Eqv) in the following sections. The first approach is to relax the binary variables into continuous ones while the second approach aims for decoupling the task assignment and wireless resource allocation.

For the first approach, first, we relax \eqref{C:binary constraint} into continuous constraints expressed as
\begin{align}
\pi(l,k)\in[0,1],\; \forall l\in\mathcal{L},\, k\in\mathcal{K}\cup\{K+1\}.\label{C:continuous constraint}
\end{align}
Therefore, the relaxed problem is expressed as:
\begin{align*}
\mathrm{(P1)}:&~\mathop{\mathtt{Minimize}}_{\mv\Pi,\{t_k^{off},t_k^{dl},t_k^c\},t_0^c}~~~t_1^{off}+t_1^c+\sum_{k=1}^Kt_k^{dl}\\
&\mathtt {Subject \ to}~~~\eqref{C:energy constraint at the source}-\eqref{C:all users assigned constraint},\, \eqref{C:nonnegative time},\, \eqref{C:min computing time constraint for helper 1}-\eqref{C:computation deadline constraint for the local user},\, \eqref{C:continuous constraint}.
\end{align*} It is worthy of noting that, since \(E_0^c\) (c.f.~\eqref{eq:local computation energy w.r.t time}) and \(E_0^{off}\) (c.f.~\eqref{eq:offloading commun. energy}) are obtained by convex operations on perspective of convex functions \((\sum_l\pi(l,K+1)C_l)^3\) and \(f(\sum_l\pi(l,k)T_l)\)'s with respect to (w.r.t.) the variables \(t_0^c\) and \(t_k^{off}\)'s, respectively, they are also convex functions. So are \(E_k^c\) and \(E_k^{dl}\), \(\forall k\in\mathcal{K}\). Therefore, (P1) is a convex problem. 
Next, we need to round the continuous \(\pi(l,k)\)'s into binary one such that \eqref{C:all tasks assigned constraint} and \eqref{C:all users assigned constraint} are satisfied. The details of the proposed joint task assignment and wireless resource allocation scheme will be discussed in Section \ref{sec:Joint Task Assignment and Wireless Resource Allocation}. In addition, we also provide a brief discussion regarding one special case of this approach in Section \ref{subsec:Fixed-Frequency Task Assignment and Wireless Resource Allocation}, in which computation frequencies of all the WDs are fixed to be their maximum,  thus serving as a benchmark scheme without computation allocation.     

For the second approach, first, it is easy to verify that given \(\mv\Pi\) fixed, (P0-Eqv) reduces to be a convex problem shown as below:
\begin{align*}
\mathrm{(P2)}:&~\mathop{\mathtt{Minimize}}_{\{t_k^{off},t_k^{dl},t_k^c\},t_0^c}~~~t_1^{off}+t_1^c+\sum_{k=1}^Kt_k^{dl}\\
&\mathtt {Subject \ to}~~~\eqref{C:energy constraint at the source}-\eqref{C:max frequency at the kth helper},\, \eqref{C:nonnegative time},\, \eqref{C:min computing time constraint for helper 1}-\eqref{C:computation deadline constraint for the local user}.
\end{align*}
Then, we decouple the design of task assignment and wireless resource allocation by employing a greedy task assignment based heuristic algorithm that will be elaborated in Section \ref{subsec:Greedy Task Assignment Based Wireless Resource Allocation}.

\section{Joint Task Assignment and Wireless Resource Allocation}\label{sec:Joint Task Assignment and Wireless Resource Allocation}
{\color{black}The main thrust of the proposed scheme in this section is to relax the binary task-assignment variables into continuous ones, and to solve the relaxed convex problem in semi-closed forms, which are then followed by attaining suboptimal task assignment based on the optimal solution to the relaxed problem.}

It is seen that problem (P1) is convex, and can thus be efficiently solved by some off-the-shelf convex optimization tools such as CVX \cite{CVX}. To gain more insights into the optimal rate and computation frequency allocation, in this section, we propose to solve (P1) leveraging the technique of Lagrangian dual decomposition. The (partial) Lagrangian of (P1) is expressed as
\begin{multline}
\mathcal{L}_1(\mv\Pi,\{t_k^{off}, t_k^{dl}, t_k^c\}, t_0^c;\eta,\beta_0,\lambda_0,\zeta_0,\mv\lambda^T, \mv\beta^T, \mv\zeta^T)=t_1^{off}+t_1^c+\sum_{k=1}^Kt_k^{dl}+\eta(\sum_{k=1}^Kt_k^{off}-t_1^{off}-t_1^c)+\beta_0\\
(t_0^c-t_1^{off}-t_1^c-\sum_{k=1}^Kt_k^{dl})
+\lambda_0\Bigg(\kappa_0\frac{\left(\sum_{l=1}^L\pi(l,K+1)C_l\right)^3}{(t_0^c)^2}+\sum_{k=1}^K\frac{1}{\bar h_k}f\left(\frac{\sum_{l=1}^L\pi(l,k)T_l}{t_k^{off}}\right)t_k^{off}- E_0\Bigg)
\\
-\zeta_0\Big(t_0^c-\frac{\sum_{l=1}^L\pi(l,K+1)C_{l}}{f_0^{\max}}\Big)
+\sum_{k=1}^K\lambda_k\Bigg(\kappa_k\frac{\left(\sum_{l=1}^L\pi(l,k)C_l\right)^3}{(t_k^c)^2}+\frac{1}{\bar g_k}f\left(\frac{\sum_{l=1}^L\pi(l,k)R_l}{t_k^{dl}}\right)t_k^{dl}\\
- E_k\Bigg)+\sum_{k=2}^K\beta_k\big(t_k^c- t_1^{off}-t_1^c-\sum_{j=1}^{k-1}t_j^{dl}+\sum_{j=1}^kt_j^{off}\big)
-\sum_{k=1}^K\zeta_k\Big(t_k^c-\frac{\sum_{l=1}^L\pi(l,k)C_{l}}{f_k^{\max}}\Big), \label{eq:Lagrangian of (P1)}
\end{multline}
where $\eta$, $\beta_0$, $\lambda_0$, and $\zeta_0$ denote the dual variables associated with the constraints \eqref{C:min computing time constraint for helper 1}, \eqref{C:computation deadline constraint for the local user}, \eqref{C:energy constraint at the source}, and \eqref{C:max frequency at the source}, respectively; \(\mv\lambda=(\lambda_1,\ldots,\lambda_k)^T\) represent the dual variables associated with the total energy constraints \eqref{C:energy constraint at the kth helper} each for one helper; \(\mv\beta=(\beta_2,\ldots,\beta_K)^T\) are the dual variables for the constraints given by \eqref{C:computation deadline constraint for helper k>=2}; and the multipliers \(\mv\zeta=(\zeta_1,\ldots,\zeta_K)^T\) are assigned to the constraints given by \eqref{C:max frequency at the kth helper}. After some manipulations, \eqref{eq:Lagrangian of (P1)} can be equivalently expressed as
\begin{multline}
\mathcal{L}_1(\mv\Pi,\{t_k^{off}, t_k^{dl}, t_k^c\}, t_0^c;\eta,\beta_0,\lambda_0,\zeta_0,\mv\lambda^T, \mv\beta^T, \mv\zeta^T)\kern-1.8pt=\kern-1.8pt\bar{\mathcal{L}}_0(\mv\Pi,t_0^c;\beta_0,\lambda_0,\zeta_0)+\zeta_0\frac{\sum_{l=1}^L\pi(l,K+1)C_{l}}{f_0^{\max}}\\
+\sum_{k=1}^K\Big(\bar{\mathcal{L}}_k(\mv\Pi,t_k^{off}, t_k^{dl}, t_k^c;\eta,\beta_0,\lambda_0,\mv\lambda^T,\mv\beta^T,\mv\zeta^T)+\zeta_k\frac{\sum_{l=1}^L\pi(l,k)C_{l}}{f_k^{\max}}\Big),\label{eq:equivalent Lagrangian of (P2)}
\end{multline}
where
\begin{align}
\bar{\mathcal{L}}_0(\mv\Pi,t_0^c;\beta_0,\lambda_0,\zeta_0)=(\beta_0-\zeta_0)t_0^c+\lambda_0\kappa_0\frac{\left(\sum_{l=1}^L\pi(l,K+1)C_l\right)^3}{(t_0^c)^2},\label{eq:bar L_0}
\end{align} and
\begin{multline}
\bar{\mathcal{L}}_k(\mv\Pi,t_k^{off}, t_k^{dl}, t_k^c;\eta,\beta_0,\lambda_0,\mv\lambda^T,\mv\beta^T,\mv\zeta^T)=A_kt_k^{dl}+B_kt_k^c+D_kt_k^{off}+\lambda_k\kappa_k\frac{\left(\sum_{l=1}^L\pi(l,k)C_l\right)^3}{(t_k^c)^2}\\
+\frac{\lambda_0}{\bar h_k}f\left(\frac{\sum_{l=1}^L\pi(l,k)T_l}{t_k^{off}}\right)t_k^{off}
+\frac{\lambda_k}{\bar g_k}f\left(\frac{\sum_{l=1}^L\pi(l,k)R_l}{t_k^{dl}}\right)t_k^{dl},\label{eq:bar L_k}
\end{multline} with \(A_k\), \(B_k\), \(D_k\), \(\forall k\in\mathcal{K}\) given by
\begin{align}
A_k=\begin{cases}
1-\beta_0-\sum_{j=k+1}^K\beta_j &k<K\\
1-\beta_0 &k=K
\end{cases},\ 
B_k=\begin{cases}
1-\eta-\beta_0-\sum_{k=2}^K\beta_k-\zeta_1 &k=1\\
\beta_k-\zeta_k &k>1
\end{cases},\label{eq:A_k and B_k}
\end{align} and
\begin{align}  
D_k=\begin{cases}
1-\beta_0 &k=1\\
\eta+\sum_{j=k}^K\beta_j &k>1
\end{cases},\label{eq:D_k}
\end{align} respectively. 
The dual function corresponding to \eqref{eq:equivalent Lagrangian of (P2)} can be expressed as
\begin{align}
g(\eta,\beta_0,\lambda_0,\zeta_0,\mv\lambda^T, \mv\beta^T, \mv\zeta^T)=\kern-4pt\min\limits_{{\eqref{C:all tasks assigned constraint}-\eqref{C:all users assigned constraint}}\atop{\eqref{C:nonnegative time}, \eqref{C:continuous constraint}}}\kern-4pt
\mathcal{L}_1(\mv\Pi,\{t_k^{off}, t_k^{dl}, t_k^c\}, t_0^c;\eta,\beta_0,\lambda_0,\zeta_0,\mv\lambda^T, \mv\beta^T, \mv\zeta^T).\label{eq:dual function of (P2)}
\end{align}  
As a result, the dual problem of \(\mathrm{(P1)}\) is formulated as
\begin{align}\mathrm{(P1\text{-}dual)}:~\mathop{\mathtt{Maximize}}_{\eta,\beta_0,\lambda_0,\zeta_0,\mv\lambda,\mv\beta,\mv\zeta}
& ~~~g(\eta,\beta_0,\lambda_0,\zeta_0,\mv\lambda^T, \mv\beta^T, \mv\zeta^T)\notag\\
\mathtt {Subject \ to}& ~~~\eta\ge 0,\, \beta_0\ge 0,\, \lambda_0\ge 0,\, \zeta_0\ge 0,\, \mv{\lambda}\ge 0,\, \mv\beta\ge 0,\, \mv\zeta\ge 0.\label{C:non-negative dual variables}
\end{align} 

\subsection{Dual-Optimal Solution to (P1)}
In this subsection, we aim for solving problem \(\mathrm{(P1\text{-}dual)}\).  To facilitate solving the optimum \(\{t_k^{off},t_k^{dl},t_k^c\}\) and \(t_0^c\) to \eqref{eq:dual function of (P2)} providing that \(\mv\Pi=\bar{\mv\Pi}\) and a set of dual variables are given, we decompose the above problem into \(K+1\) subproblems including $K$ for \(\forall k\in\mathcal{K}\) and one for \(k=K+1\) as follows.
\begin{align*}
\mathrm{(P1\text{-}sub1)}:~\mathop{\mathtt{Minimize}}_{t_k^{off},t_k^{dl},t_k^c}
&~~~\bar{\mathcal{L}}_k(\bar{\mv\Pi},t_k^{off},                                                                       t_k^{dl},t_k^c;\eta,\beta_0,\lambda_0,\mv\lambda^T,\mv\beta^T,\mv\zeta^T)\\
\mathtt {Subject \ to}& ~~~t_k^{off}\ge 0,\, t_k^{dl}\ge 0.\\
\mathrm{(P1\text{-}sub2)}:~\mathop{\mathtt{Minimize}}_{t_0^c}
&~~~\bar{\mathcal{L}}_0(\bar{\mv\Pi},t_0^c;\beta_0,\lambda_0,\zeta_0).
\end{align*} 
Since these problems are independent of each other, they can be solved in parallel each for one $k$, \(k\in\mathcal{K}\cup\{K+1\}\). 

Next, define \(\tilde f(x)\triangleq \frac{B}{\ln 2}(W_0(\frac{x}{e}-\frac{1}{e})+1)\) for \(x\ge 0\), in which \(W_0(\cdot)\) is the principal branch of Lambert $W$ function defined as the inverse function of \(xe^x=y\) \cite{corless96LambertW}. Then, in accordance with the optimal solution to the above subproblems, the optimal time and power together with the optimal task assignment to \eqref{eq:dual function of (P2)}  are shown in the following proposition.    
\begin{proposition}
Given a set of dual variables, the optimal solution to \eqref{eq:dual function of (P2)} is given by
\begin{align}
&\hat t_k^{off}=\begin{cases}
\frac{\sum_{l=1}^L\hat\pi(l,k)T_l}{\tilde f\left(\Myfrac{D_k\bar h_k}{\lambda_0}\right)} &\mbox{if}\ D_k>0,\\
\inf & \mbox{otherwise};
\end{cases}\ 
\hat t_k^{dl}=\begin{cases}
\frac{\sum_{l=1}^L\hat\pi(l,k)R_l}{\tilde f\left(\Myfrac{A_k\bar g_k}{\lambda_k}\right)} &\mbox{if}\ A_k>0,\\
\inf & \mbox{otherwise};
\end{cases}\label{eq:optimum transmission time given pi}\\
&\hat t_k^c=\begin{cases}
\frac{\sum_{l=1}^L\hat\pi(l,k)C_l}{\sqrt[\leftroot{-2}\uproot{1}3]{\Myfrac{B_k}{(2\lambda_k\kappa_k)}}} &\mbox{if}\ B_k>0,\\
\inf & \mbox{otherwise};
\end{cases}\ \mbox{and}\ \
\hat t_0^c=\begin{cases}
\frac{\sum_{l=1}^L\hat\pi(l,K+1)C_l}{\sqrt[\leftroot{-2}\uproot{1}3]{\Myfrac{(\beta_0-\zeta_0)}{(2\lambda_0\kappa_0)}}} &\mbox{if}\ \beta_0-\zeta_0>0,\\
\inf & \mbox{otherwise}.
\end{cases}\label{eq:optimum execution time given pi}
\end{align} 
In addition, \(\hat\pi(l,k)\)'s shown in \eqref{eq:optimum transmission time given pi} and \eqref{eq:optimum execution time given pi} denote the optimal solution to the following linear programming (LP) problem:
\begin{align*}
\mathrm{(LP1)}:~\mathop{\mathtt{Minimize}}_{\mv\Pi}
&~~~\sum_{l=1}^L\left(\sum_{k=1}^K\phi_{l,k}\pi(l,k)+\phi_{l,K+1}\pi(l,K+1)\right)\\
\mathtt {Subject \ to}& ~~~\eqref{C:all tasks assigned constraint}-\eqref{C:all users assigned constraint},\, \eqref{C:continuous constraint},
\end{align*} where \(\phi_{l,k}\), \(\forall k\in\mathcal{K}\), \(\forall l\in\mathcal{L}\), is given by
\begin{multline}
\phi_{l,k}=\frac{A_kR_l}{\tilde f\left(\Myfrac{A_k\bar g_k}{\lambda_k}\right)}+\frac{B_kC_l}{\sqrt[\leftroot{-2}\uproot{1}3]{\Myfrac{B_k}{(2\lambda_k\kappa_k)}}}+\frac{D_kT_l}{\tilde f\left(\Myfrac{D_k\bar h_k}{\lambda_0}\right)}+\lambda_k\kappa_kC_l\left(\frac{B_k}{2\lambda_k\kappa_k}\right)^{\frac{2}{3}}\\
+\frac{\lambda_0}{\bar h_k}f\left(\tilde f\left(\frac{D_k\bar h_k}{\lambda_0}\right)\right)\frac{T_l}{\tilde f\left(\Myfrac{D_k\bar h_k}{\lambda_0}\right)}+\frac{\lambda_k}{\bar g_k}f\left(\tilde f\left(\frac{A_k\bar g_k}{\lambda_k}\right)\right)\frac{R_l}{\tilde f\left(\Myfrac{A_k\bar g_k}{\lambda_k}\right)}+\zeta_k\frac{C_l}{f_k^{\max}},\label{eq:phi_{l,k}}
\end{multline} and \(\phi_{l,K+1}\), \(\forall l\in\mathcal{L}\), is expressed as
\begin{align}
\phi_{l,K+1}=\frac{(\beta_0-\zeta_0)C_l}{\sqrt[\leftroot{-2}\uproot{1}3]{\Myfrac{(\beta_0-\zeta_0)}{(2\lambda_0\kappa_0)}}}+\lambda_0\kappa_0C_l\left(\frac{\beta_0-\zeta_0}{2\lambda_0\kappa_0}\right)^{\frac{2}{3}}+\zeta_0\frac{C_l}{f_0^{\max}}.\label{eq:phi_l,K+1}
\end{align}
\label{prop:solution to the dual function}
\end{proposition}
\begin{IEEEproof}
Please refer to Appendix~\ref{appendix:proof of solution to the dual function}. 
\end{IEEEproof}
Accordingly, problem (P1-dual) can be further modified as shown below.
\begin{align}
\mathrm{(P1\text{-}dual^\prime)}:~\mathop{\mathtt{Maximize}}_{\eta,\beta_0,\lambda_0,\zeta_0,\mv\lambda,\mv\beta,\mv\zeta}
& ~~~g(\eta,\beta_0,\lambda_0,\zeta_0,\mv\lambda^T, \mv\beta^T, \mv\zeta^T)\notag\\
\mathtt {Subject \ to}& ~~~\eqref{C:non-negative dual variables},\notag\\
& ~~~A_k\ge 0,\ B_k\ge 0,\ D_k\ge 0,\; \forall k\in\mathcal{K},\ \beta_0-\zeta_0\ge 0,\label{C:dual problem constraints}.
\end{align} 

\begin{remark}
Note that some useful insights can be drawn from the results in Proposition~\ref{prop:solution to the dual function}. First, with the dual variables given, \(\tilde f(\Myfrac{D_k\bar h_k}{\lambda_0})\) and \(\tilde f(\Myfrac{A_k\bar g_k}{\lambda_k})\) can be, respectively, interpreted (in terms of the dual function \eqref{eq:dual function of (P2)}) as the optimum offloading rate to helper $k$ and the optimum results downloading rate from helper $k$, while  \(\sqrt[\leftroot{-2}\uproot{1}3]{\Myfrac{B_k}{(2\lambda_k\kappa_k)}}\) and \(\sqrt[\leftroot{-2}\uproot{1}3]{\Myfrac{(\beta_0-\zeta_0)}{(2\lambda_0\kappa_0)}}\) represent the optimum computation frequencies at the $k$th helper and the local user, respectively. Accordingly, when helper $k$ enjoys good offloading (downloading) channel gain, the optimum offloading (downloading ) rate \(\tilde f(\Myfrac{D_k\bar h_k}{\lambda_0})\) (\(\tilde f(\Myfrac{A_k\bar g_k}{\lambda_k})\)) also gets large due to non-decreasing monotonicity of \(W_0(x)\). Moreover, provided that the total energy constraint for the local user is violated, thereby incurring a larger Lagrangian multiplier \(\lambda_0\) (c.f.~\eqref{eq:Lagrangian of (P1)}), the optimum offloading time (\(\hat t_k^{off}\)'s) and the optimum local computation time (\(\hat t_0^c\)) under the same \(\mv\Pi\) turn out to be longer. Hence, the total energy consumption for the local user gets reduced, which complies with Lemma~\ref{lemma:decreasing function of the communications energy}. %Similarly, the maximum computation frequency relates to the solution given by \eqref{eq:optimum execution time given pi} in the sense that, e.g., if the maximum computation frequency constraint for user $k$ is violated with an intermediate \(\zeta_k\) sufficiently large, \(B_k\) (c.f.~\eqref{eq:A_k and B_k}) gets relatively small and thus longer execution time \(\hat t_k^c\) (c.f.~\eqref{eq:optimum execution time given pi}) is required.  
\end{remark}

In a sum, given an initial (feasible) set of dual variables, the optimal solution to \eqref{eq:dual function of (P2)} is first obtained leveraging Proposition~\ref{prop:solution to the dual function}, and then the dual variables are readily updated utilizing some sub-gradient based method, e.g., ellipsoid method \cite{EE364b}, until a predefined threshold controlling the accuracy of the algorithm is achieved.

\subsection{Primal-Optimal Solution to (P1)}
We aim for solving (P1) in this subsection. Since there might exist multiple solutions to (LP1) in each iteration of the ellipsoid method while there is only one optimal solution to the convex problem (P1), we retrieve the primal-optimal \(\mv\Pi\) to (P1) through the dual-optimal solution in the sequel. Denoting the optimum dual variables by \((\eta^\ast,\beta_0^\ast,\lambda_0^\ast,\zeta_0^\ast,\mv\lambda^{\ast T},\mv\beta^{\ast T},\mv\zeta^{\ast T})\), the primal-optimal solution are related to the dual-optimal one as follows (c.f.~\eqref{eq:optimum transmission time given pi} and \eqref{eq:optimum execution time given pi}).
\begin{align}
\frac{\sum_{l=1}^L\pi(l,k)T_l}{t_k^{off}} =\tilde f\left(\Myfrac{D_k^\ast\bar h_k}{\lambda_0^\ast}\right),\ \ \ \ \ \
& \frac{\sum_{l=1}^L\pi(l,k)R_l}{t_k^{dl}}=\tilde f\left(\Myfrac{A_k^\ast\bar g_k}{\lambda_k^\ast}\right),\label{eq:primal-dual transmission rate}\\ 
\frac{\sum_{l=1}^L\pi(l,k)C_l}{t_k^c} =\sqrt[\leftroot{-2}\uproot{1}3]{\Myfrac{B_k^\ast}{(2\lambda_k^\ast\kappa_k)}},\ \ \ \ \ \
&\frac{\sum_{l=1}^L\pi(l,K+1)C_l}{t_0^c} =\sqrt[\leftroot{-2}\uproot{1}3]{\Myfrac{(\beta_0^\ast-\zeta_0^\ast)}{(2\lambda_0^\ast\kappa_0)}},\label{eq:primal-dual computation frequency}
\end{align} 
where \(A_k^\ast\)'s, \(B_k^\ast\)'s, and \(D_k^\ast\)'s are obtained by plugging the optimum dual variables into \eqref{eq:A_k and B_k} and \eqref{eq:D_k}. Then transform problem (P1) w.r.t. variables \(\mv\Pi\), \(\{t_k^{off},t_k^{dl},t_k^c\}\), and \(t_0^c\) into an LP w.r.t. \(\mv\Pi\) by plugging \eqref{eq:primal-dual transmission rate} and \eqref{eq:primal-dual computation frequency} into (P1). Denoting this LP by (LP2), (LP2) is then readily solved by standard LP algorithm, e.g., simplex method\footnote{\color{black}The simplex method is a standard search algorithm that travels through the set of basic feasible solutions one at a time, until the optimal basic feasible solution (if it exists) is identified \cite{Murty1983LP}, e.g., {\em linprog} in Matlab.}.

Note that \eqref{C:max frequency at the source} and \eqref{C:max frequency at the kth helper} should be satisfied with \(\sqrt[\leftroot{-2}\uproot{1}3]{\tfrac{\beta_0^\ast-\zeta_0^\ast}{(2\lambda_0^\ast\kappa_0)}}\le f_0^{\max}\) and \(\sqrt[\leftroot{-2}\uproot{1}3]{\tfrac{B_k^\ast}{(2\lambda_k^\ast\kappa_k)}}\le f_k^{\max}\), respectively, independent of \(\mv\Pi\), and thus can be safely removed from the constraints of (LP2). Denoting the optimal solution to (LP2) by \(\mv\Pi^\ast\), the primal-optimal solution to (P1) are thus given by
\begin{align}
t_k^{off\ast}  =
\frac{\sum_{l=1}^L\pi^\ast(l,k)T_l}{\tilde f\left(\Myfrac{D_k^\ast\bar h_k}{\lambda_0^\ast}\right)},\ \ \ \ \ \  
& t_k^{dl\ast} =
\frac{\sum_{l=1}^L\pi^\ast(l,k)R_l}{\tilde f\left(\Myfrac{A_k^\ast\bar g_k}{\lambda_k^\ast}\right)},\label{eq:optimum transmission time}\\ 
t_k^{c\ast}  =
\frac{\sum_{l=1}^L\pi^\ast(l,k)C_l}{\sqrt[\leftroot{-2}\uproot{1}3]{\Myfrac{B_k^\ast}{(2\lambda_k^\ast\kappa_k)}}},\ \ \ \ \ \ 
& t_0^{c\ast} =
\frac{\sum_{l=1}^L\pi^\ast(l,K+1)C_l}{\sqrt[\leftroot{-2}\uproot{1}3]{\Myfrac{(\beta_0^\ast-\zeta_0^\ast)}{(2\lambda_0^\ast\kappa_0)}}}.\label{eq:optimum computation time}
\end{align} 
 
\subsection{Sub-Optimal Solution to (P0-Eqv)}
In this section, we propose a suboptimal scheme to jointly optimize task assignment as well as  time and power allocation for (P0-Eqv) based on the optimal solution to (P1) developed in the previous subsections. 

\begin{figure}[htp]
	\centering
	\includegraphics[height=5.8in]{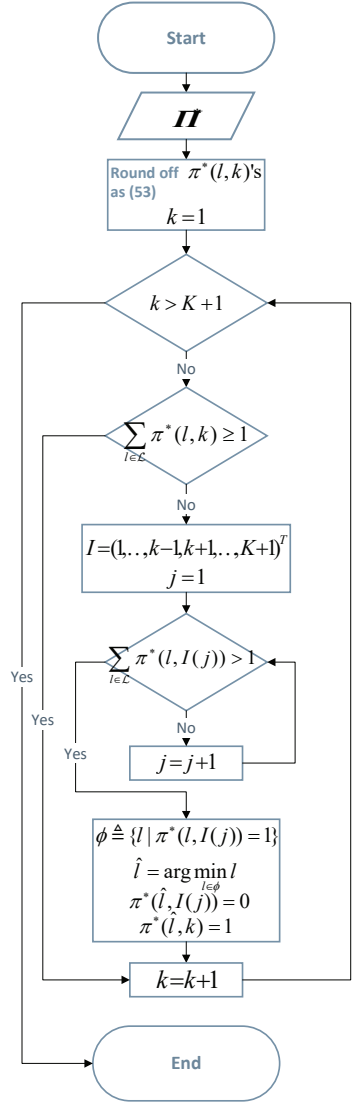}
	\caption{The procedures to modify \(\mv\Pi^\ast\) such that the constraints in \eqref{C:all tasks assigned constraint}-\eqref{C:binary constraint} are satisfied.}\label{fig:round-ff}
	\vspace{-0.3in}
\end{figure}

First, we propose to round off \(\pi^\ast(l,k)\)'s as follows such that \eqref{C:binary constraint} are satisfied: 
\begin{align}
\pi^\ast(l,k)=\begin{cases}
	1 &\mbox{if}\ k=\hat k_l,\\
	0 & \mbox{otherwise},
\end{cases}\; \forall l\in\mathcal{L},\label{eq:intermediate modified Pi}
\end{align} where \(\hat k_l=\arg\kern-4pt\max\limits_{k\in\mathcal{K}\cup\{K+1\}}\kern-4pt\pi^\ast(l,k)\) {\color{black}To ensure that each helper is assigned at least one task, we need to further adjust \(\mv\Pi^\ast\) to avoid the cases where some WD is assigned with no task after rounding off as \eqref{eq:intermediate modified Pi}.} The detailed procedure of constructing such \(\mv\Pi\) is shown in Fig.~\ref{fig:round-ff}.

Next, given the updated \(\mv\Pi^\ast\), what remains to be solved reduces to problem (P2). Hence, it can  be solved using Lagrangian dual decomposition following similar procedures as shown in Section~\ref{sec:Joint Task Assignment and Wireless Resource Allocation}.~A. The proposed joint task assignment and wireless resource allocation scheme  is thus summarized in Algorithm~\ref{alg:Algorithm I}.
\begin{algorithm}[htp]
	\caption{The Proposed Suboptimal Algorithm for (P0-Eqv)}\label{alg:Algorithm I}
	{\bf Input} a set of dual variables satisfying \eqref{C:non-negative dual variables} and \eqref{C:dual problem constraints}: \(\eta^{(0)},\beta_0^{(0)},\lambda_0^{(0)},\zeta_0^{(0)},\mv\lambda^{(0)},\mv\beta^{(0)}, \mv\zeta^{(0)}\)
	\begin{enumerate}
		\item Solve \(\mathrm{(P1\text{-}dual^\prime)}\) using ellipsoid method \cite{EE364b} to obtain the dual-optimal solution: \(\eta^\ast,\beta_0^\ast,\zeta_0^\ast,\mv\lambda^\ast,\mv\beta^\ast,\mv\zeta^\ast\);
		\item Solve (LP2) using simplex method to retrieve the primal-optimal \(\mv\Pi^\ast\); 
%		\item Obtain the primal-optimal \(\{t_k^{off\ast},t_k^{dl\ast},t_k^{c\ast}\}\) and \(t_0^{c\ast}\) to (P1) in accordance with \eqref{eq:optimum transmission time} and \eqref{eq:optimum computation time};
		\item Round off \(\pi^\ast_{l,k}\)'s in accordance with \eqref{eq:intermediate modified Pi};
		\item Modify  \(\mv\Pi^\ast\) in accordance with the procedures shown in Fig. \ref{fig:round-ff};
		\item Solve (P2) given the modified \(\mv\Pi^\ast\).
	\end{enumerate}
	{\bf Output} solution to (P0-Eqv) 
\end{algorithm}

\subsection{Complexity}
The complexity of Algorithm~\ref{alg:Algorithm I} mainly lies in solving (P1). Hence, we focus on discussing the complexity of solving (P1). Specifically, the complexity of solving (P1) includes solving \(\mathrm{(P1\text{-}dual^\prime)}\) by ellipsoid method via primal-dual iterations, and solving (LP1) by simplex method in each iteration of the ellipsoid method. In accordance with the worst-case number of iterations for the ellipsoid method and the expected complexity of the simplex algorithm\footnote{\color{black}Despite that the worst-case complexity of simplex algorithm is known to be exponential, it was shown in \cite{Spielman2004simplex} that most LP can be approximated with their inputs perturbed and then solved by simplex algorithm in polynomial time.}, the complexity of Algorithm~\ref{alg:Algorithm I} can be estimated by  
\begin{align}
\mathcal{O}\left(18(K+1)^2\log(\Myfrac{\sqrt{\gamma}H}{\epsilon})\mathcal{P}(L(K+1))\right),\label{eq:complexity of Algorithm I}
\end{align}
where \(H\triangleq\max\limits_{\mv h\in\partial g(\mv z),\, \mv z\in E^{(0)}}\|\mv h\|\) is a Lipschitz constant for \eqref{eq:dual function of (P2)} over the initial ellipsoid \(E^{(0)}=\{\mv z\left.\vert\right. \|\mv z\|\le\sqrt{\gamma}\}\) \cite[Ellipsoid Method (notes)]{EE364b}, \(\mv h\) is a sub-gradient of \(g(\eta,\beta_0,\lambda_0,\zeta_0,\mv\lambda^T, \mv\beta^T, \mv\zeta^T)\) over \(E^{(0)}\), and \(\epsilon\) is a parameter controlling the accuracy of the algorithm. 

\section{Low-Complexity Benchmark Schemes}\label{sec:Benchmark Schemes}
In this section, we develop two low-complexity benchmark schemes, which are provided in Sections \ref{subsec:Fixed-Frequency Task Assignment and Wireless Resource Allocation} and \ref{subsec:Greedy Task Assignment Based Wireless Resource Allocation}, respectively. 

{\color{black}\subsection{Fixed-Frequency Task Assignment and Wireless Resource Allocation}\label{subsec:Fixed-Frequency Task Assignment and Wireless Resource Allocation}}
{\color{black}In this subsection, we consider a benchmark scheme that alleviates the WDs from adjusting their computation frequencies using DVFS by endowing them with the maximum computation capacities, i.e., \(f_0=f_0^{\max}\), and \(f_k=f_k^{\max}\), \(k\in\mathcal{K}\). Note that with the computation frequencies fixed, this scheme reduces to a special case of the joint task assignment and wireless resource allocation scheme with the constraints \eqref{C:max frequency at the source} and \eqref{C:max frequency at the kth helper} being active. Since we have studied a similar fixed-frequency design in \cite{xing2018fixed}, in the sequel, we only focus on some results that are distinguished from Section \ref{sec:Joint Task Assignment and Wireless Resource Allocation} due to space limitation.}  

{\color{black}First, with the computation frequencies fixed, the computation energy consumed by the local user and the helper turn out to be \(E_0^c=\kappa_0\sum\limits_{l\in\mathcal{L}}\pi(l,K+1)C_l(f_0^{\max})^2\), and \(E_k^c=\kappa_k\sum\limits_{l\in\mathcal{L}}\pi(l,k)C_l(f_k^{\max})^2\), respectively.  Moreover, \(t_0^c\) and \(t_k^c\)'s in problem (P0) are safely removed by replacing them with the left-hand side (LHS) of \eqref{C:max frequency at the source} and \eqref{C:max frequency at the kth helper}, respectively. In addition, since \(t_1^c\) cannot be adjusted once helper $1$ is assigned with its tasks, \(I_1\) (c.f.~\eqref{eq:waiting time of helper 1}) cannot be further simplified. To sum up, problem (P0-Eqv) reduces to be:
\begin{subequations}
\begin{align}
\mathrm{(P0^\prime\text{-}{\rm Eqv})}:&~\mathop{\mathtt{Minimize}}_{\mv\Pi,\{t_k^{off},t_k^{dl}\}}~~~I_1+\sum_{k=1}^Kt_k^{dl}\notag\\
&~~~\mathtt {Subject \ to}\ \ \eqref{C:energy constraint at the source}-\eqref{C:energy constraint at the kth helper},\; \eqref{C:all tasks assigned constraint}-\eqref{C:nonnegative time},\notag\\
&~~~\sum_{k=1}^Kt_k^{off}\le I_1, \\
&~~~t_1^{off}+\frac{\sum_{l=1}^L\pi(l,1)C_{l}}{f_1^{\max}}\le I_1, \\
&~~~\frac{\sum_{l=1}^L\pi(l,k)C_{l}}{f_k^{\max}}\le I_1+\sum_{j=1}^{k-1}t_j^{dl}-\sum_{j=1}^kt_j^{off}, \; \forall k\in\mathcal{K}\setminus \{1\}, \\
&~~~\frac{\sum_{l=1}^L\pi(l,K+1)C_{l}}{f_0^{\max}}\le I_1+\sum_{k=1}^Kt_k^{dl}. 
\end{align}
\end{subequations}}

{\color{black}\begin{remark}
Compared with problem (P0-Eqv), problem \(\mathrm{(P0^\prime\text{-}{\rm Eqv})}\) is more sceptical to infeasibility\footnote{Discussion regarding the feasibility of the fixed-frequency scheme can be referred to \cite{xing2018fixed}.}, since the minimum of \(E_0^c\)  (\(E_k^c\)'s) cannot reach zero even with \(t_0^c\) (\(t_k^c\)'s) approaching to infinity (c.f.~\eqref{eq:local computation energy w.r.t time} (\eqref{eq:remote computation energy})). Once it is feasible, the solution to \(\mathrm{(P0^\prime\text{-}{\rm Eqv})}\) can be found by a simplified version of Algorithm~\ref{alg:Algorithm I} that yields an approximate complexity of \(\mathcal{O}(2(2K+3)^2\log(\Myfrac{\sqrt{\gamma}H}{\epsilon})\mathcal{P}(L(K+1)))\), which is lower than \eqref{eq:complexity of Algorithm I} due to decrease in the number of dual variables.  
\end{remark}}

\subsection{Greedy Task Assignment Based Wireless Resource Allocation}\label{subsec:Greedy Task Assignment Based Wireless Resource Allocation}
The joint task assignment and wireless resource allocation and its special case with fixed computation frequencies both admit complexity involving a polynomial factor \(\mathcal{P}(L(K+1))\)due to solving \(\hat\pi(l,k)\)'s using simplex method. To further reduce computation complexity, we propose in this subsection greedy task assignment based wireless resource allocation. {\color{black}The main idea of this heuristic scheme is as follows. First, we obtain the objective value of problem (P2) as a cost function of the initial task assignment. Then we assign one task each time from those unallocated to a WD who yields the least amount of increase in the cost function. Next, since two different task selection criteria are applied, we compare the results of these two sub-schemes and  choose the one that achieves the lower total latency.} 
 
%Specifically, each sub-scheme comprises four major steps as follows. 1) Sort the tasks with the quantity of one feature (input data length $T_l$'s/output data length $R_l$'s) in ascending order as a sequence. 2) Assign the first task in the present task sequence to the unassigned helper with the best channel condition (\(\bar h_k\)'s/\(\bar g_k\)'s), and then remove this task from the sequence until the first $K$ tasks are assigned. 3) Assign the first task in the present sequence to the user that yields the minimum increase among all the users in the objective value of (P2), and then remove this task from the sequence until all of the $L$ tasks are assigned. 4) With the task assignment matrix obtained as above, solve (P2) to find the optimal rate and frequency allocation.
The greedy task assignment based wireless resource allocation scheme is shown in Algorithm~\ref{alg:Algorithm II}. In Algorithm~\ref{alg:Algorithm II}, it is seen that for the first $K$ tasks, each of them is in turn assigned to the helper with the best channel condition among those who have not yet been occupied. The intuition behind such assignment is that the selected helper will consume the least amount of energy in transmission (c.f.~\eqref{eq:offloading commun. energy} and \eqref{eq:downloading commun. energy}), and thus any spare energy can be exploited for further latency reduction. It is also worth noting that the task with the longest (input/output) data flow is executed locally for the sake of saving data-transmission time.
  
\begin{algorithm}[!htp]
	\caption{The Heuristic Algorithm for (P0-Eqv)}\label{alg:Algorithm II}
	{\bf Input} \(T_l\), \(R_l\), \(l\in\mathcal{L}\)
	\begin{enumerate}
		\item Initialize \(\mv\Pi^{(0)}=\mv 0_{L\times(K+1)}\);\label{step:start}
		\item Sort \(T_l\)'s in ascending order such as \(T_{m_1}\le\ldots\le T_{m_L}\);
		\item \(\pi^{(0)}(m_L,K+1)=1\), \(\mathcal{L}^{(0)}=\mathcal{L}\setminus\{m_L\}\), \(\mathcal{K}^{(0)}=\mathcal{K}\setminus\{K+1\}\), \(i=0\);
		\item {\bf Repeat}
		\item \(i=i+1\), \(\mv\Pi^{i}=\mv\Pi^{(i-1)}\);
		\item \(\pi^{(i)}(m_i,k^\ast)\)=1, where \(k^\ast=\arg\max\limits_{k\in\mathcal{K}^{(i-1)}}\bar h_k\);
		\item \(\mathcal{L}^{(i)}=\mathcal{L}^{(i-1)}\setminus\{m_i\}\), \(\mathcal{K}^{(i)}=\mathcal{K}^{(i-1)}\setminus\{k^\ast\}\);
		\item {\bf Until} \(i=K\); 
		\item Solve (P2) with \(\mv\Pi\)  given by \(\mv\Pi^{(i)}\) and obtain the objective value \(p^{(0)}\);
		\item {\bf Repeat}
		\item \(i=i+1\), \(j=0\);
		\item {\bf Repeat}
		\item \(j=j+1\), \(\mv\Pi^{(j)}=\mv\Pi^{(i-1)}\);
		\item \(\pi^{(j)}(m_i,j)=1\);
		\item Solve (P2) with \(\mv\Pi\) given by \(\mv\Pi^{(j)}\) and obtain the objective value \(p^{(j)}\);
		\item {\bf Until} \(j=K+1\);
		\item \(\pi^{(i)}(m_i,j^\ast)=1\), where \(j^\ast=\arg\min\limits_{j\in\mathcal{K}\cup\{K+1\}}p^{(j)}\);
		\item \(\mathcal{L}^{(i)}=\mathcal{L}^{(i-1)}\setminus\{m_i\}\);
		\item {\bf Until} \(\mathcal{L}^{(i)}=\emptyset\);
		\item Solve (P2) with \(\mv\Pi\) given by \(\mv\Pi^{(i)}\) and obtain the objective value \(p_1(\mv\Pi,\{t_k^{off},t_k^{dl},t_k^c\},t_0^c)\);\label{step:end}
		\item Repeat steps \ref{step:start})-\ref{step:end}) with \(T_{m_l}\)'s, \(\bar h_{n_k}\)'s, and \(p_1(\mv\Pi,\{t_k^{off},t_k^{dl},t_k^c\},t_0^c)\) replaced by \(R_{m_l}\)'s, \(\bar g_{n_k}\)'s, and \(p_2(\mv\Pi,\{t_k^{off},t_k^{dl},t_k^c\},t_0^c)\), respectively;
		\item \(n^\ast=\arg\min\limits_{n\in\{1,2\}}p_n(\mv\Pi,\{t_k^{off},t_k^{dl},t_k^c\},t_0^c)\).	
	\end{enumerate}
		{\bf Output} \(\arg p_{n^\ast}(\mv\Pi,\{t_k^{off},t_k^{dl},t_k^c\},t_0^c)\) as solution to (P0-Eqv)
\end{algorithm}  

The complexity of Algorithm~\ref{alg:Algorithm II} comprises that of two sub-schemes (with sorting \(\{T_l\}\) and \(\{R_l\}\), respectively), each of which requires solving problem (P2) for \((L-K-1)(K+1)\) times to find the right task assignment matrix. Since solving (P2) by ellipsoid method yields maximum as many as \(18(K+1)^2\log(\Myfrac{\sqrt{\gamma}H}{\epsilon})\) iterations (c.f.~\eqref{eq:complexity of Algorithm I}), the worst-case complexity of Algorithm~\ref{alg:Algorithm II} is given by
\begin{align}
\mathcal{O}\left(36(L-K-1)(K+1)^3\log(\Myfrac{\sqrt{\gamma}H}{\epsilon})\right).\label{eq:complexity of Algorithm II}
\end{align}  
In fact, \eqref{eq:complexity of Algorithm II} suggests that even the worst-case complexity of Algorithm~\ref{alg:Algorithm II} is much less than that of Algorithm~\ref{alg:Algorithm I} (c.f.~\eqref{eq:complexity of Algorithm I}) as long as \(2(L-K-1)(K+1)<\mathcal{P}(L(K+1))\), which is easily satisfied in most cases.

\section{Numerical Results}\label{sec:Numerical Results}
In this section, we provide numerical results to validate the effectiveness of the proposed joint task assignment and wireless resource allocation ({\bf Joint optimization}) in Section \ref{sec:Joint Task Assignment and Wireless Resource Allocation}, as compared against the fixed-frequency scheme ({\bf Fixed frequency}) and the greedy task assignment based algorithm ({\bf Greedy assignment}) presented in Section \ref{sec:Benchmark Schemes} as well as other benchmark schemes as follows. 
\begin{itemize}
\item {\bf Optimal} The MINLP problem (P0-Eqv) is solved by exhaustive search over feasible \(\mv\Pi\) for \((K+1)^L-\sum_{i=1}^K(-1)^{i+1}\binom{K+1}{i}(K+1-i)^L\) times\footnote{This number take all combinations of \(\mv\Pi\) satisfying \eqref{C:all tasks assigned constraint}-\eqref{C:binary constraint} into account.} with a convex problem (P2) solved each time for a given \(\mv\Pi\). Note that ``Optimal'' is of exponential complexity and is thus too costly to implement in practice. Hence, we only provide this scheme for one numerical example in the sequel. 
%\item {\bf Fixed frequency} The local (helper's) computation frequency \(f_0\) (\(f_k\)'s) is fixed by its maximum limit \(f_0^{\max}\) (\(f_k^{\max}\)'s), and therefore the corresponding computation energy is given by \(E_0^c=\kappa_0\sum_{l\in\mathcal{L}}\pi(l,K+1)C_l(f_0^{\max})^2\) (\(E_k^c=\kappa_k\sum_{l\in\mathcal{L}}\pi(l,k)C_l(f_k^{\max})^2\)). With \(E_0^c\) and \(E_k\)'s replaced by these modified ones and the constraints in \eqref{C:max frequency at the source}-\eqref{C:max frequency at the kth helper} removed, (P0-Eqv) is solved under fixed frequency of the local user and the $K$ helpers. This scheme in fact serves as a special case of the proposed scheme in Section~\ref{sec:Joint Task Assignment and Wireless Resource Allocation}, and therefore similar procedure as Algorithm~\ref{alg:Algorithm I} can be applied under fixed frequency, the details of which are omitted herein for brevity. 
\item {\bf Random assignment} In this scheme, a random matrix with its entries drawn from $i.i.d.$ uniform distribution over $[0,1]$ is first generated, and then the procedure shown in Fig.~\ref{fig:round-ff} is employed to construct a feasible \(\mv\Pi\) to (P0-Eqv). Next, solve problem (P2) under this \(\mv\Pi\). 
\item {\bf Local execution} All of the computation tasks in $\mathcal{L}$ are executed locally with the total latency expressed as \(\max\left\{\sqrt{\frac{\kappa_0(\sum_{l\in\mathcal{L}}C_l)^3}{E_0}},\frac{\sum_{l\in\mathcal{L}}C_l}{f_0^{\max}}\right\}\) by combining \eqref{C:energy constraint at the source} and \eqref{C:max frequency at the source}. 
\end{itemize}

In simulations, the $K$ helpers are located with their distance uniformly distributed over \([0,500]\)m away from the local user. The wireless channel model consists of pathloss and Rayleigh fading. The distance-dependent pathloss model is given by \(128.1+37.6\log 10(d)\) in dB, where \(d\) in km is the distance between the local user and a helper. The Rayleigh fading is generated by $i.i.d.$ CSCG RVs with zero mean and unit variance. The capacitance coefficients are set all equal as \(\kappa_k=\kappa_0=10^{-28}\), \(\forall k\in\mathcal{K}\) \cite{You2017offloading}. We also assume identical AWGN power over the transmission bandwidth  of $B=312.5$KHz and the noise power spectrum density of $-169$ dBm/Hz. The other parameters are set as follows unless otherwise specified. The bit-length of the input and output data are set as \(T_l\sim\mathcal{U}[0,10^4]\) bits and \(R_l\sim\mathcal{U}[0,10^4]\) bits. The amount of computation required per task is assumed to be \(C_l\sim\mathcal{U}[0,5\times10^6]\) cycles. The energy constraints are set as \(E_0=-30\)dB and \(E_k=-20\)dB, \(\forall k\in\mathcal{K}\). The maximum frequency for local computing is \(f_0^{\max}=.9\)GHz, and that for remote computing is \(f_k^{\max}\sim\mathcal{U}[1.5,2]\)GHz, \(\forall k\in\mathcal{K}\). The total latency is obtained by averaging over $300$ times of channel realizations.

\subsection{The Effect of Wireless Resource on the Total Latency}
\begin{figure}[htp]
\centering
\includegraphics[width=3.3in]{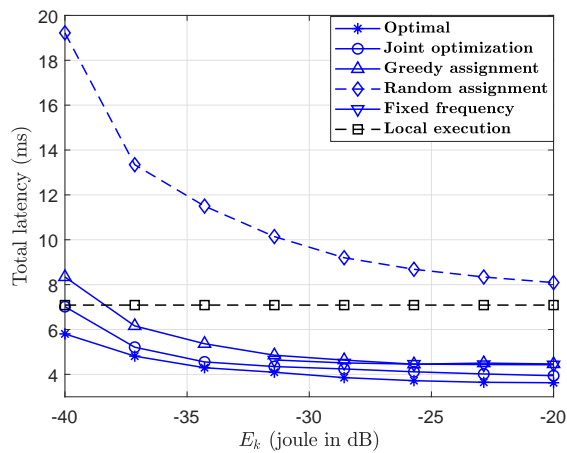}
\caption{The total latency versus the energy constraints with $K=2$ and $L=5$.}\label{fig:gap with the optimal}
\end{figure}
We consider a simple scenario where the local user has $L=5$ tasks to be executed in the present of $K=2$ helpers. Fig.~\ref{fig:gap with the optimal} shows the total latency versus the energy constraints at the helpers assuming $E_1=E_2$. It is observed that the optimal scheme outperforms all the other ones, while our proposed ``Joint optimization'' achieves the second lowest total latency with little gap to the optimal solution. ``Greedy assignment'' admits reducing total latency with larger helpers' energy constraints, and outperforms ``Local execution'' in most cases except for those with $E_k$'s below around $-38$dB, since when the helpers also suffer from a scarcity of energy, full local execution with constant total latency of \(\max\left\{\sqrt{\frac{\kappa_0(\sum_{l\in\mathcal{L}}C_l)^3}{E_0}},\frac{\sum_{l\in\mathcal{L}}C_l}{f_0^{\max}}\right\}\) is intuitively better than computation offloading. Moreover, ``Fixed frequency'' does not work until  $E_k$'s is larger than $-32$dB due to the infeasibility of  (P0-Eqv).

\begin{figure}[htp]
	\centering
	\includegraphics[width=3.3in]{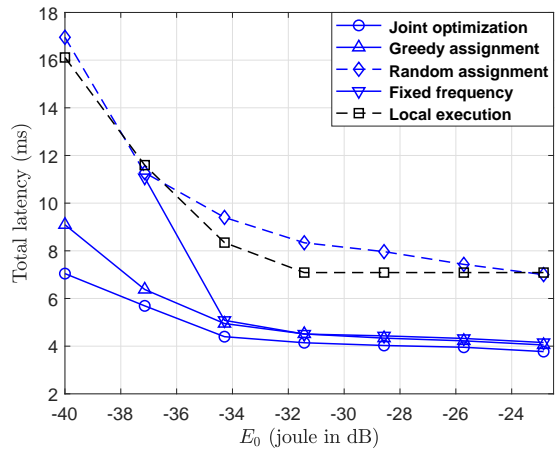}
	\caption{The total latency versus the energy constraints at the local user with $E_1=E_2=-10$dB and $L=5$.}\label{fig:total latency vs local energy}
\end{figure}
Considering the same scenario as in Fig.~\ref{fig:gap with the optimal}, Fig.~\ref{fig:total latency vs local energy} shows the total latency versus the energy constraint at the local user with $E_1=E_2=-10$dB. It is seen that the average total latency decreases over $E_0$ for all of the schemes. In particular, ``Joint optimization'', ``Greedy assignment'' and ``Fixed frequency'' admit sharp reduce in the average total latency when $E_0$ is less than about $-34$dB, and slightly go down when $E_0$ continues decreasing. This complies with intuition as follows. When the energy resource is scarce at the local user, a little more energy will cause significant decrease in the computation offloading time, while as for $E_0$ beyond $-34$dB, the bottleneck mainly lies in the helper' energy constraint $E_k$'s. It is also worth noting that ``Fixed frequency'' is substantially outperformed by ``Joint optimization'' and  ``Greedy assignment'' when $E_0$ is less than about $-34$dB, since local computing with its full computation capacity is far from optimality under circumstances of limited energy supply. In addition, for ``Local execution'',  \(\frac{\sum_{l\in\mathcal{L}}C_l}{f_0^{\max}}\) starts taking effect when $E_0$ is larger than about $-31.5$dB.   

\begin{figure}[htp]
	\centering
	\includegraphics[width=3.3in]{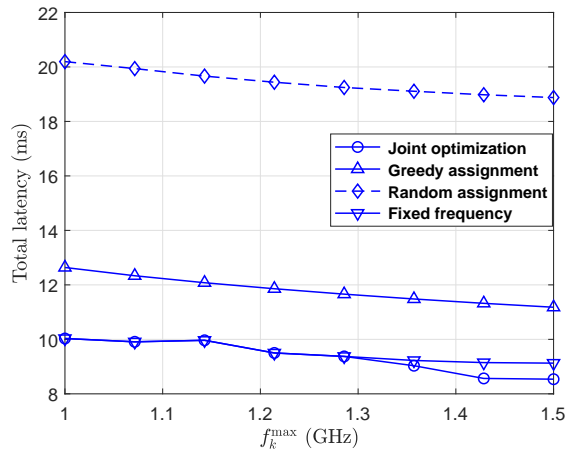}
	\caption{The total latency versus the maximum remote computation frequency with $K=5$ and $L=7$.}\label{fig:total latency vs helper frequency}
\end{figure}
Fig.~\ref{fig:total latency vs helper frequency} shows the total latency versus the maximum computation frequencies at the helpers under the assumption of \(f_1^{\max}=\ldots=f_5^{\max}\) and $L=7$. {\color{black}``Fixed frequency'' is observed to almost overlap with ``Joint optimization'' in most cases (when \(f_k^{\max}\)'s is below about $1.32$GHz), and to outperform ``Greedy assignment''. This is because  when \(f_k^{\max}\)'s is below \(1.32\)GHz, the helpers' optimal computation frequencies obtained by ``Joint optimization'' tend to be their maximum capacities, i.e., \(f_k^{\max}\)'s, which is almost equivalent to ``Fixed frequency'' (c.f.~Section \ref{subsec:Fixed-Frequency Task Assignment and Wireless Resource Allocation}).} However, when \(f_k^{\max}\)'s continuously increases, ``Joint optimization'' will suppress the helpers' computation capacities strictly below their limits such that under the given helper's energy budget, there is still sufficient energy left for results downloading thus achieving the overall better performance. In addition, ``Local execution'' under this setting is equal to $39.5$ms, which is too large to illustrate and is thus removed from the figure herein. 

\subsection{The Effect of Computation Load on the Average Total Latency}

\begin{figure}[htp]
	\begin{subfigure}[t]{.49\textwidth}
	\centering
	\includegraphics[width=\linewidth]{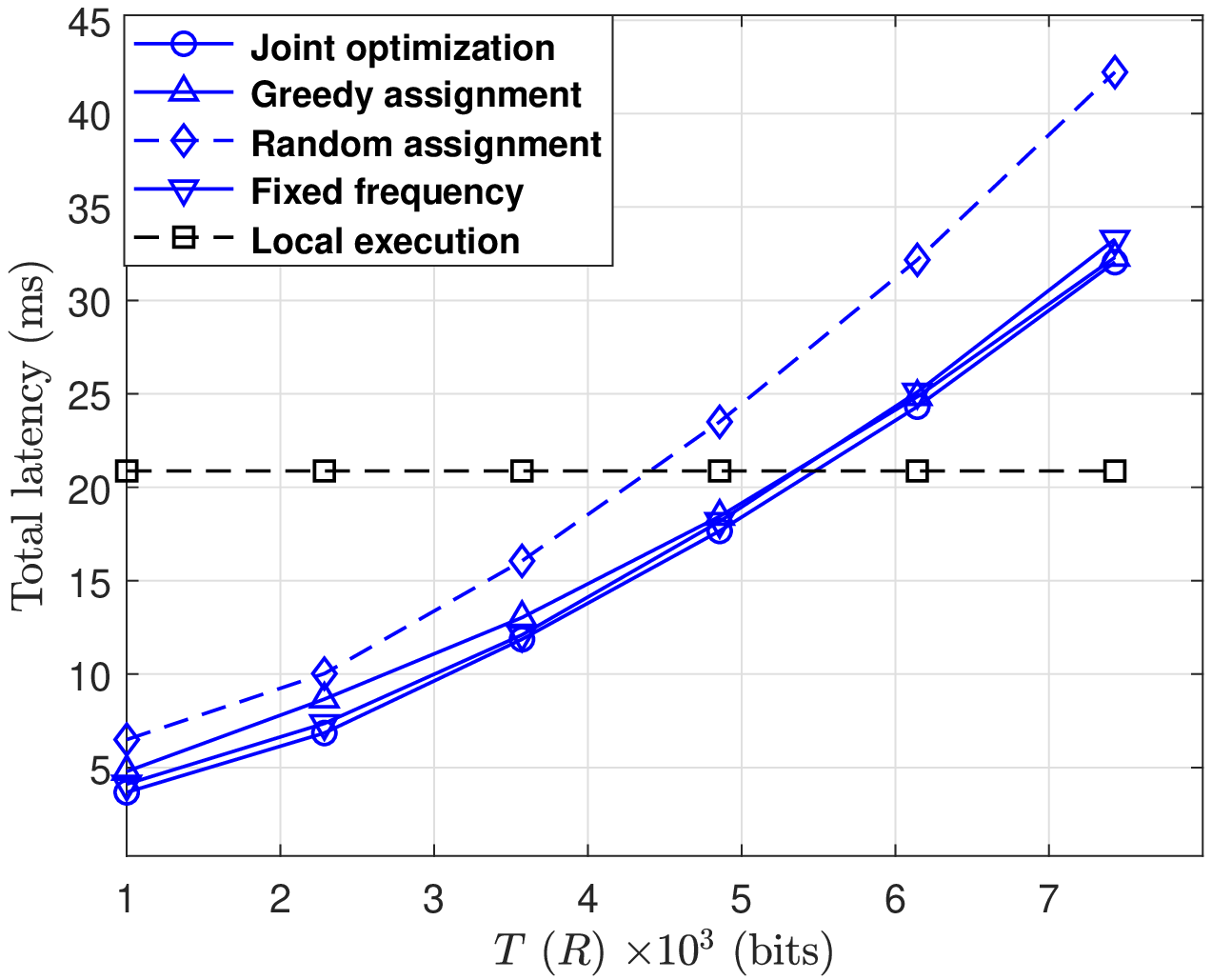}
	\caption{\(E_0=-30\)dB, \(E_1=\ldots=E_5=-10\)dB, and $L=8$.}\label{subfig:subfig:total latency vs data size low energy}
	\end{subfigure}
	\begin{subfigure}[t]{.49\textwidth}
	\centering
	\includegraphics[width=\linewidth]{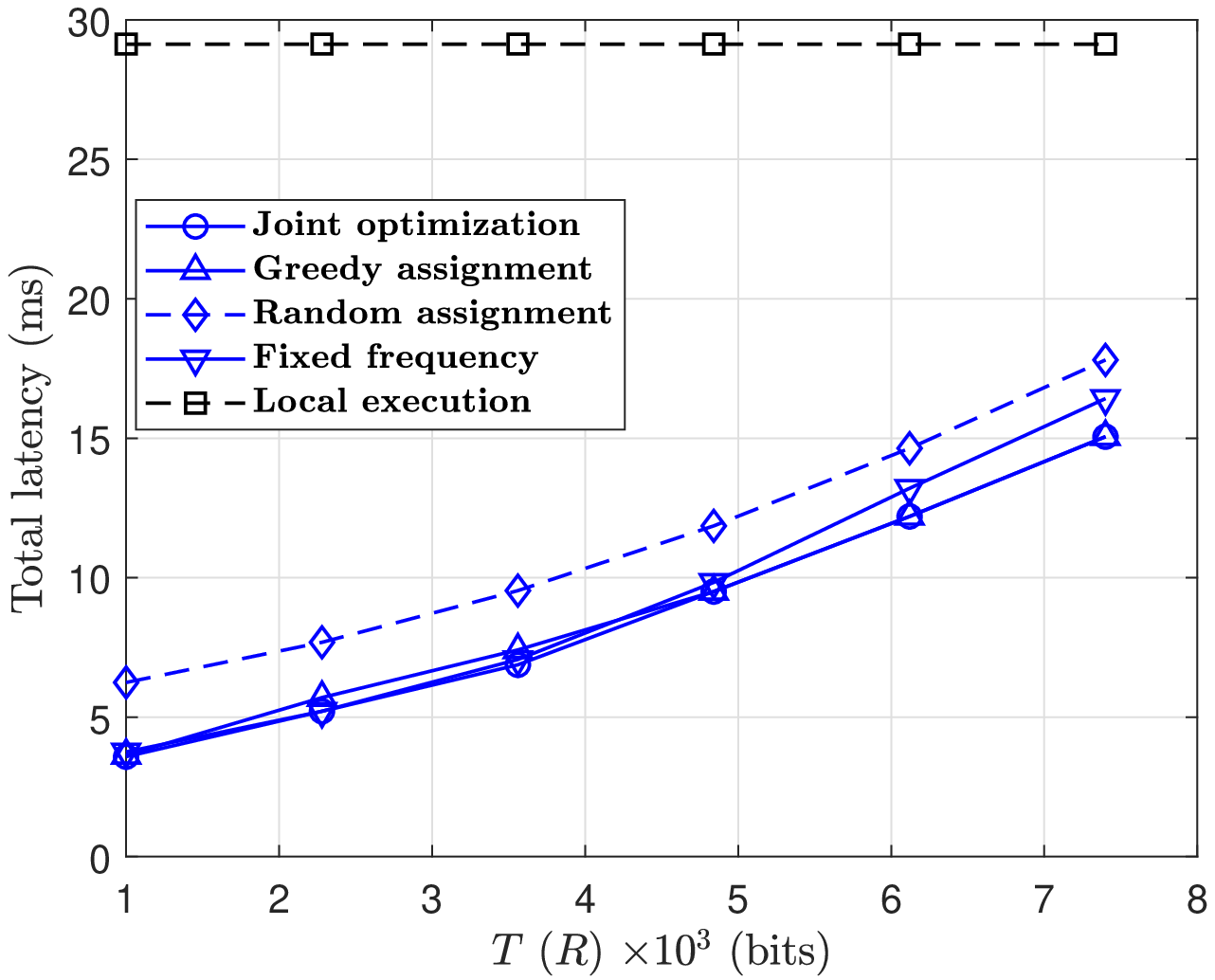}
	\caption{\color{black}\(E_0=-33\)dB, \(E_1=\ldots=E_5=0\)dB, and $L=7$.}\label{subfig:total latency vs data size high energy}
    \end{subfigure}
\caption{The total latency versus the bit-length of the input (output) task data.}\label{fig:total latency vs data size}
\vspace{-.2in}	
\end{figure}
 
The trade-offs between the total latency and the bit-length of the task input/output data are demonstrated in Fig.~\ref{fig:total latency vs data size} with different energy constraints under the setup of equal input/output length across all tasks, i.e., \(T_1=T\) and \(R_1=R\), \(\forall l\in\mathcal{L}\). It is observed that the schemes of ``Joint optimization'', ``Fixed frequency'', and ``Greedy assignment'' almost overlap each other. Among these three schemes, ``Greedy assignment'' is slightly worse than the other two in the lower range of data length in Fig.~(\ref{subfig:subfig:total latency vs data size low energy}), while ``Fixed frequency'' is noticeably worse than the other two in the higher range of data length in Fig.~(\ref{subfig:total latency vs data size high energy}). ``Random assignment'' continues being outperformed by the above three task offloading schemes especially when $T$ ($R$) gets larger, which validates the importance of effective task assignment. 
{\color{black}It is also worth noting that with the energy setting of \(E_0=-30\)dB at the local user and \(E_k=-10\)dB, \(\forall k\in\mathcal{K}\), in the helpers, when $T$ ($R$) is larger than around $5.5\times 10^3$~bits, ``Local execution'' becomes favourable due to increasing energy consumption in data transmission. By contrast, when the energy budget of the helpers increases to \(0\)dB in Fig.~(\ref{subfig:total latency vs data size high energy}), all the task offloading schemes considerably outperform ``Local execution'', since with sufficient energy supply (c.f.~\eqref{C:energy constraint at the kth helper}), it costs the helpers less amount of time to transmit even very long task-output data.}      

\begin{figure}[htp]
	\begin{subfigure}[t][][t]{.49\textwidth}
			\centering
			\resizebox{.95\linewidth}{!}{%
				\begin{tabular}{@{}c|llllllll@{}}
					\toprule
					C (\(\times 10^6\)cycles) & 1.00 & 2.29 & 3.57 & 4.86 & 6.14 & 7.43 & 8.71 & 10.0 \\ \midrule
					 Local execution latency (ms) & 7.77 & 20.2 & 39.5 & 62.7 & 89.2 & 119 & 151 & 185 \\ \bottomrule
				\end{tabular}%
			}
		\includegraphics[width=\linewidth]{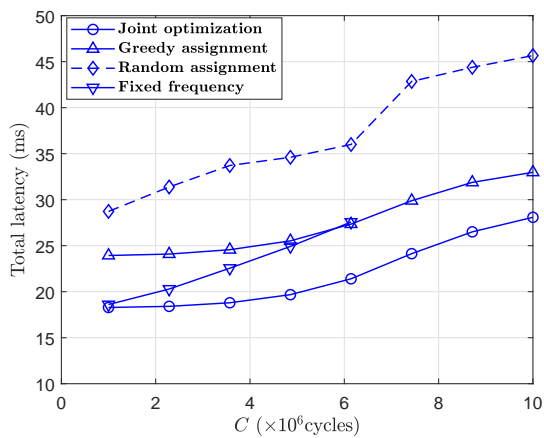}\caption{\(E_0=-30\)dB, \(E_1=\ldots=E_5=-20\)dB, and $L=7$.}\label{subfig:total latency vs computing load low energy}
	\end{subfigure}
	\hfill
	\begin{subfigure}[t][][t]{.49\textwidth}
		\centering
			\resizebox{.95\linewidth}{!}{%
				\begin{tabular}{@{}c|llllllll@{}}
					\toprule
					C (\(\times 10^6\)cycles) & 1.00 & 2.29 & 3.57 & 4.86 & 6.14 & 7.43 & 8.71 & 10.0 \\ \midrule
					Local execution latency (ms) & 14.1 & 48.9 & 95.5 & 151 & 215 & 286 & 364 & 447 \\ \bottomrule
				\end{tabular}%
			}
		\includegraphics[width=\linewidth]{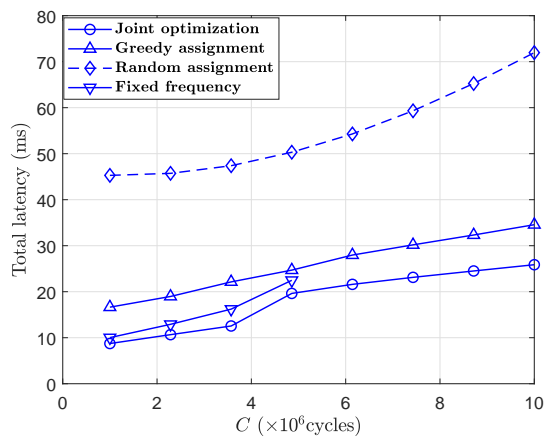}\caption{\color{black}\(E_0=-33\)dB, \(E_1=\ldots=E_6=0\)dB, and $L=10$.}\label{subfig:total latency vs computing load high energy}
	\end{subfigure}
\caption{The total latency versus the amount of computation required per task.}\label{fig:total latency vs computing load}
\vspace{-.2in}
\end{figure}

The effect of the amount of computation required per task on the total latency is shown in Fig.~\ref{fig:total latency vs computing load} with different energy constraints' setup assuming \(C_l=C\), \(\forall l\in\mathcal{L}\).\footnote{Due to the relatively high regime into which the total latency of ``Local execution'' falls in Fig.~\ref{fig:total latency vs computing load}, we present the results in tables to improve readability.} It is seen that ``Fixed frequency'' turns out to be unavailable 
when $C$ is larger than about $7.4\times 10^6$ cycles and $6.0\times 10^6$ cycles in Figs.~(\ref{subfig:total latency vs computing load low energy}) and (\ref{subfig:total latency vs computing load high energy}), respectively, due to infeasibility incurred for the same reason as discussed in Fig.~\ref{fig:gap with the optimal}. It is also observed that the schemes of ``Joint optimization'', ``Greedy assignment'' and ``Random assignment'' increase with $C$ very slowly  when it is below about $5.0\times 10^6$cycles ($4.0\times 10^6$cycles) in Fig.~(\ref{subfig:total latency vs computing load low energy}) (Fig.~(\ref{subfig:total latency vs computing load high energy})), which is because under low to medium computation load per task, very little increase in \(t_1^c\) is already sufficient to satisfy \eqref{C:computation deadline constraint for helper k>=2}. However, when $C$ gets larger, the increasing amount of computation energy yields decrease in communications energy thus prolonging the total latency. 
{\color{black}In addition, ``Local execution'' admits the best performance among all the schemes when the computation burden is  below $3.1\times 10^6$~cycles per task under the lower energy setting of the helpers in Fig.~(\ref{subfig:total latency vs computing load low energy}), since ``Local execution'' is preferred for computation non-intensive tasks. However, when there is drastic difference in the energy budget between the local user and the helpers, the advantage of the task assignment schemes is prominently observed in Fig.~(\ref{subfig:total latency vs computing load high energy}), as similarly explained for Fig.~(\ref{subfig:total latency vs data size high energy}).}   

\begin{figure}[htp]
	\centering
	\includegraphics[width=3.3in]{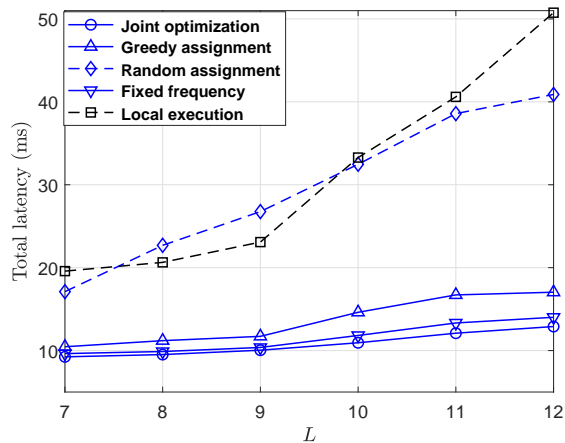}
	\caption{The total latency versus the number of tasks with $K=5$.}\label{fig:total latency vs number of tasks}
\end{figure}
Fig.~\ref{fig:total latency vs number of tasks} shows the total latency under different number of tasks with $K=5$ helpers. It is seen that while the average total latency achieved by ``Joint optimization'', ``Greedy assignment'', and ``Fixed frequency'' steadily increase with the total number of tasks, that of ``Local execution'' grows fast, which motivates cooperative MEC especially when there are a large number of tasks to be executed. Moreover, the performance of ``Random assignment'', almost worst among all the schemes, stresses the importance of proper task distribution schemes. ``Greedy assignment'' is seen to strike satisfying balance between the performance and the complexity.

\section{Concluding Remarks}\label{sec:Conclusion}
In this paper, we investigated the joint task assignment, communications rate, as well as computation frequency allocation for a D2D-enabled multi-helper MEC system assuming binary task offloading. Under a TDMA communication protocol, we aimed for minimizing the overall computation latency subject to individual energy and computation capacity constraints at both the local user and the helpers. Since the formulated problem was an MINLP that is in general difficult to solve, we proposed an efficient convex-relaxation-based algorithm to construct a suboptimal task assignment solution based on the optimal one to the relaxed problem. Furthermore, a benchmark scheme with fixed computation frequency and a greedy task assignment based heuristic algorithm were also developed to strike the balance between complexity and performance. Finally, numerical results verified that our proposed design is an effective solution to enhance the local user's computation latency, by exploiting D2D collaborations at the network edge. 

{\color{black}Due to space limitation, there are several other challenging issues yet to be addressed in this paper, which will be investigated in our future work. First, this paper assumed that all of the WDs are deployed at fixed locations with static wireless channels. In practice, since these WDs may move over time, the wireless channels will fluctuate over time, and the D2D connections established between the local user and the helpers may also be dropped as they move away from each other. Under these circumstances,  we need to consider new design principles (e.g., online algorithms with long-term objectives capturing dynamics of the D2D links) to combat such mobility issues \cite{Chen16D2D}. Second, in this paper we considered TDMA owing to its easy implementation in practice. Other orthogonal multiple access schemes, e.g., OFDMA \cite{You2017offloading,Tran2017MEC}, and more sophisticated NOMA schemes, e.g., sparse code multiple access (SCMA) \cite{Ding17survey}, can be employed to further enhance the system performance. Under these schemes, on top of mixed-integer task assignment, how to design subcarrier allocation for OFDMA and joint message decoding for NOMA are also quite challenging issues worthy of further study. In addition, our considered model assumed that the helpers have agreed to cooperate in the computation offloading, while we believe there are also different types of incentive driven collaboration schemes that require sophisticated design. At last, it is also worth investigating how to extend our current single-user multi-helper MEC model to a multi-user multi-helper one by proper design of multi-user scheduling algorithms.}       

\appendices
\section{Proof of Proposition~\ref{prop:simplification of T^total}}\label{appendix:proof of simplification of T^total}
First, in order to prove Proposition~\ref{prop:simplification of T^total}, we need the following lemma.
\begin{lemma}
	The function \(h(y,t)=f\left(\frac{y}{t}\right)t\) monotonically decreases over \(t>0\).\label{lemma:decreasing function of the communications energy}
\end{lemma}
\begin{IEEEproof}
The monotonicity can be obtained by evaluating the first-order partial derivative of $h(y,t)$ w.r.t. $t$, and using the fact that \((1-x)e^x-1<0\), for \(x>0\).
\end{IEEEproof}

On one hand, there are two possible cases for the optimal \(I_k\)'s given by \eqref{eq:waiting time of helper k>=2}: case 1) \(\sum_{j=1}^kt_j^{off}+t_k^c>I_{k-1}+t_{k-1}^{dl}\); and case 2) \(\sum_{j=1}^kt_j^{off}+t_k^c\le I_{k-1}+t_{k-1}^{dl}\). In line with Lemma~\ref{lemma:decreasing function of the communications energy}, the total transmitting energy of the $k$th helper, i.e., \(E_k^{dl}\)'s (c.f.~\eqref{eq:downloading commun. energy}), \(\forall k\in\mathcal{K}\), monotonically decreases over \(t_k^{dl}\)'s. Hence, if the first case occurs, helper $k-1$ (\(k\ge 2\)) can slow down its downloading, i.e., extending \(t_{k-1}^{dl}\), until \(\sum_{j=1}^kt_j^{off}+t_k^c=I_{k-1}+t_{k-1}^{dl}\) is satisfied (c.f.~Fig. \ref{fig:frame protocol}), such that $I_k$ remains unchanged but the transmitting energy of helper $k-1$ gets reduced. As such, the two cases can be, w.l.o.g., merged into one as \(\sum_{j=1}^kt_j^{off}+t_k^c\le I_{k-1}+t_{k-1}^{dl}\), \(\forall k\in\mathcal{K}\setminus \{1\}\), which suggests the following constraint on \(t_k^c\) given by
\begin{align}
t_k^c\le I_{k-1}+t_{k-1}^{dl}-\sum_{j=1}^kt_j^{off}. \label{eq:recursive computation deadline constraint for helper k>=2}
\end{align}
As it is seen clear that \eqref{eq:waiting time of helper k>=2} reduces to
\begin{align}
I_k=I_{k-1}+t_{k-1}^{dl}, \; \forall k\in\mathcal{K}\setminus \{1\}, \label{eq:recursive I_k}
\end{align}
the waiting time for helper $k$ (\(k\ge 2\)) can be recursively obtained as
\begin{align}
I_k=I_1+\sum_{j=1}^{k-1}t_j^{dl}. \label{eq:waiting time expression of helper k>=2}
\end{align}

On the other hand, there are also two possible cases for the optimal $I_1$ given by \eqref{eq:waiting time of helper 1}: case 1) \(t_1^{off}+t_1^c<\sum_{k=1}^Kt_k^{off}\); and case 2) \(t_1^{off}+t_1^c\ge \sum_{k=1}^Kt_k^{off}\). Since the $k$th helper's computation energy $E_k^c$ given by \eqref{eq:remote computation energy}, \(\forall k\in\mathcal{K}\), monotonically decreases over \(t_k^c\)'s, if case 1) takes place, it is always possible for helper $1$ to slow down its computation such that \(t_1^{off}+t_1^c=\sum_{k=1}^Kt_k^{off}\) is met without violating the other constraints. In a sum, $I_1$, w.l.o.g., reduces to 
\begin{align}
I_1=t_1^{off}+t_1^c, \label{eq:waiting time expression of helper 1}
\end{align} subject to \eqref{C:min computing time constraint for helper 1}. Combining \eqref{eq:waiting time expression of helper k>=2} with \eqref{eq:waiting time expression of helper 1}, and substituting the results for \(I_{k-1}\) in \eqref{eq:recursive computation deadline constraint for helper k>=2}, \eqref{C:computation deadline constraint for helper k>=2} follows. 

Furthermore, substituting \eqref{eq:waiting time expression of helper 1} for $I_1$ in \eqref{eq:waiting time expression of helper k>=2} with $k=K$, it is easy to obtain that \(I_K=t_1^{off}+t_1^c+\sum_{j=1}^{K-1}t_j^{dl}\), and thus the completion time $T$ reduces to \(T=t_1^{off}+t_1^c+\sum_jt_j^{dl}\). As a result, the total latency given by \eqref{eq:total latency} turns out to be 
\begin{align}	
T^{\rm total}=\max\{t_0^c,t_1^{off}+t_1^c+\sum_{k=1}^Kt_k^{dl}\}. \label{eq:intermediate total latency}
\end{align}
In addition, it can also be verified that when the optimal \(T^{\rm total}\) given by \eqref{eq:intermediate total latency} yields \(t_0^{c}>t_1^{off}+t_1^c+\sum_{k=1}^Kt_k^{dl}\), it is always possible for one of the $K$ helpers to slow down its downloading rate such that \(t_0^{c}=t_1^{off}+t_1^c+\sum_{k=1}^Kt_k^{dl}\) without violating the other constraints. Therefore, $T^{\rm total}$, w.l.o.g., can be further simplified as
\begin{align}
T^{\rm total}=t_1^{off}+t_1^c+\sum_{k=1}^Kt_k^{dl}, \label{eq:simplified total latency}
\end{align} subject to \eqref{C:computation deadline constraint for the local user}.

\section{Proof of Proposition~\ref{prop:solution to the dual function}}\label{appendix:proof of solution to the dual function}
First, given a set of dual variables, we solve (P2-sub1) and (P2-sub2) leveraging some of the Karush-Kuhn-Tucker (KKT) conditions as follows.
\begin{subequations}
	\begin{align}
	& D_k+\frac{\lambda_0}{\bar h_k}\left(f\left(\frac{\sum_{l=1}^L\bar\pi(l,k)T_l}{t_k^{off}}\right)-\frac{\sum_{l=1}^L\bar\pi(l,k)T_l}{t_k^{off}}f^\prime\left(\frac{\sum_{l=1}^L\bar\pi(l,k)T_l}{t_k^{off}}\right)\right)=0,\label{eq:KKT regarding t_k^off}\\
	& A_k+\frac{\lambda_k}{\bar g_k}\left(f\left(\frac{\sum_{l=1}^L\bar\pi(l,k)R_l}{t_k^{dl}}\right)-\frac{\sum_{l=1}^L\bar\pi(l,k)R_l}{t_k^{dl}}f^\prime\left(\frac{\sum_{l=1}^L\bar\pi(l,k)R_l}{t_k^{dl}}\right)\right)=0,\label{eq:KKT regarding t_k^dl}\\
	& B_k-2\lambda_k\kappa_k\left(\frac{\sum_{l=1}^L\bar\pi(l,k)C_l}{t_k^c}\right)^3=0,\label{eq:KKT regarding t_k^c}\\
	& \beta_0-\zeta_0-2\lambda_0\kappa_0\left(\frac{\sum_{l=1}^L\bar\pi(l,K+1)C_l}{t_0^c}\right)^3=0.\label{eq:KKT regarding t_0^c}
	\end{align}
\end{subequations}
Since the solution to the equation \(f(x)-xf^\prime(x)=y\) for \(x>0\) is shown to be \(x=\tilde f(-y)\) \cite{corless96LambertW}, it is easy to verify that  \eqref{eq:KKT regarding t_k^off} and \eqref{eq:KKT regarding t_k^dl} imply \eqref{eq:optimum transmission time given pi} with \(\bar\pi(l,k)\) replaced by \(\hat\pi(l,k)\) therein. Similarly, \eqref{eq:optimum execution time given pi} follows as a result \eqref{eq:KKT regarding t_k^c} and \eqref{eq:KKT regarding t_0^c}. Note that to keep the objective function of problem (P2) from being infeasible, \eqref{eq:KKT regarding t_k^off}-\eqref{eq:KKT regarding t_0^c} suggest that \(D_k>0\), \(A_k>0\), \(B_k>0\), \(\forall k\in\mathcal{K}\), and \(\beta_0-\zeta_0>0\), respectively, which also complies with the domain of the principal branch of Lambert $W$ function.

Next, to find an optimum solution to \eqref{eq:dual function of (P2)} in terms of \(\mv\Pi\) we substitute \(\frac{\sum_{l=1}^L\bar\pi(l,K+1)C_l}{\sqrt[\leftroot{-2}\uproot{1}3]{\Myfrac{(\beta_0-\zeta_0)}{(2\lambda_0\kappa_0)}}}\) for \(t_0^c\) in \eqref{eq:bar L_0}, and \(\frac{\sum_{l=1}^L\bar\pi(l,k)T_l}{\tilde f\left(\Myfrac{D_k\bar h_k}{\lambda_0}\right)}\), \(\frac{\sum_{l=1}^L\bar\pi(l,k)R_l}{\tilde f\left(\Myfrac{A_k\bar g_k}{\lambda_k}\right)}\), and \(\frac{\sum_{l=1}^L\bar\pi(l,k)C_l}{\sqrt[\leftroot{-2}\uproot{1}3]{\Myfrac{B_k}{(2\lambda_k\kappa_k)}}}\) for \(t_k^{off}\), \(t_k^{dl}\), and \(t_k^c\), respectively, in \eqref{eq:bar L_k}, such that \eqref{eq:equivalent Lagrangian of (P2)} is expressed in terms of \(\bar\pi(l,k)\)'s. Furthermore, considering \(\bar\pi(l,k)\)'s as the only (primal) variables, after some manipulations, the minimization of \eqref{eq:equivalent Lagrangian of (P2)} subject to \eqref{C:all tasks assigned constraint}, \eqref{C:all users assigned constraint}, and \eqref{C:continuous constraint} is formulated as problem (LP1), which can be solved using simplex method. Denoting the optimal solution to (LP1) as \(\hat\pi(l,k)\)'s, the results given by \eqref{eq:optimum transmission time given pi} and \eqref{eq:optimum execution time given pi} are thus obtained, which completes the proof.

\linespread{1.3}
%the map from commands to actual font sizes can be found at https://en.wikibooks.org/wiki/LaTeX/Fonts
%\balance
\bibliographystyle{IEEEtran}
%\IEEEtriggeratref{15} % This fires a new column at the given BibTeX reference number
\bibliography{MEC_ref}

\end{document}